\documentclass[a4paper,fleqn]{cas-dc}
\usepackage[numbers,sort&compress]{natbib}
\usepackage{helvet}
\IfFileExists{stix.sty}{}{\usepackage{mathptmx}}
\IfFileExists{inconsolata.sty}{}{\usepackage{courier}}
\hypersetup{hidelinks}
\newcommand{\doi}[1]{doi: \href{https://doi.org/#1}{\nolinkurl{#1}}}

\def\Tc{$T_{\rm{c}}$}

\def\T1T{$T_1T$}
\def\NÈel{N\'eel}
\def\Hc{$H_{\rm{c}}$~}
\def\Hs{$H_{\rm{s}}$}
\newcommand{\jld}{\ensuremath{J_\text{1D}}}

\newcommand{\jEd}{\ensuremath{J_\text{3D}}}
\newcommand{\Bl}{\ensuremath{B_\text{c1}}}
\newcommand{\Bu}{\ensuremath{B_\text{c2}}}

\begin{document}
\ExplSyntaxOn
\keys_set:nn { stm / mktitle } { nologo = true }
\ExplSyntaxOff
\shorttitle{UHF NMR for CondMat}
\shortauthors{R. Stern et~al.}
\title [mode = title]{Nuclear magnetic resonance in ultra high magnetic fields: Recent applications to quantum spin systems}
\author[1,2]{Raivo Stern}[type=author,
                        orcid=0000-0002-6724-9834]
\cormark[1]
\ead{raivo.stern@kbfi.ee}
\affiliation[1]{organization={National Institute of Chemical Physics and Biophysics},
                addressline={Akadeemia tee 23},
                city={Tallinn},
                postcode={12618},
                country={Estonia}}
\affiliation[2]{organization={National High Magnetic Field Laboratory},
                addressline={1800 E Paul Dirac Dr.},
                postcode={32310},
                city={Tallahassee},
                state={FL},
                country={USA}}
\author[2]{Rong Cong}[type=author,
                         orcid=0000-0002-2535-0926]
\ead{rcong@magnet.fsu.edu}
\author[1]{Ivo Heinmaa}[type=author,
                         orcid=0000-0003-4796-642X]
\ead{ivo.heinmaa@kbfi.ee}
\author[3]{Mladen Horvati\'c}[type=author,
                         orcid=0000-0001-7161-0488]
\ead{mladen.horvatic@lncmi.cnrs.fr}
\affiliation[3]{organization={Laboratoire National des Champs Magn\'{e}tiques Intenses, LNCMI-CNRS (UPR3228), EMFL, Univ. Grenoble Alpes, Univ. Toulouse, INSA-T},
                city={Grenoble},
                postcode={38042},
                country={France}}
\author[4]{Yoshihiko Ihara}[type=author,
                         orcid=0000-0003-2332-4342]
\ead{yihara@phys.sci.hokudai.ac.jp}
\affiliation[4]{organization={Department of Physics, Faculty of Science, Hokkaido University},
                city={Sapporo},
                postcode={060-0810},
                country={Japan}}
\author[3]{Steffen Kr\"{a}mer}[type=author,
                         orcid=0000-0002-6107-3583]
\ead{steffen.kramer@lncmi.cnrs.fr}
\author[5]{Hannes K\"{u}hne}[type=author,
                         orcid=0000-0002-0090-2288]
\ead{h.kuehne@hzdr.de}
\affiliation[5]{organization={Hochfeld-Magnetlabor Dresden (HLD-EMFL), Helmholtz-Zentrum Dresden-Rossendorf},
                city={Dresden},
                postcode={01328},
                country={Germany}}
\author[2]{Arneil Reyes}[type=author,
                         orcid=0000-0003-2638-5474]
\ead{reyes@magnet.fsu.edu}
\cortext[cor1]{Corresponding author}
\nonumnote{Published in \textit{Solid State Nuclear Magnetic Resonance} \textbf{145} (2026) 102127.
\enspace\href{https://doi.org/10.1016/j.ssnmr.2026.102127}{doi:10.1016/j.ssnmr.2026.102127}.}

\begin{abstract}
In this review, we present recent applications of nuclear magnetic resonance (NMR) for probing, at the microscopic level, new quantum states of condensed matter that emerge in strong static and pulsed magnetic fields. We focus on experiments performed in resistive magnets (up to 34~T) and hybrid magnets (max 45~T), further extended to fields beyond 60~T using pulsed magnets, which together offer broad access to these quantum states and the quantum phase transitions between them. After a brief overview of NMR observables, we discuss several topics: quantum spin systems (including spin chains, spin ladders, spin-nematic phases), and their exotic phenomena (magnetization plateaux, Bose--Einstein condensation of triplet excitations), as well as aspects of novel multiferroic systems.
\end{abstract}

\begin{keywords}
NMR spectroscopy \sep Magnetic resonance \sep Solid-state physics
\sep Quantum magnetism \sep Ultra-high magnetic fields \sep Pulsed magnetic fields
\end{keywords}

\maketitle
\thispagestyle{plain}
\pagestyle{plain}
\tableofcontents

\section{\label{intro}Introduction}
 Discovered in the mid-1940s by Purcell, Torrey, and Pound at Harvard~\cite{Purcell1946}, and by Bloch, Hansen, and Packard at Stanford ~\cite{Bloch1946}, nuclear magnetic resonance (NMR) is today a well-established, versatile, and widely applicable spectroscopic method across various scientific disciplines.
This progress stands on three equally important foundations: the development of superconducting (SC) magnets providing extremely stable and homogeneous (up to $10^{-10}$) magnetic fields beyond 28.2~T (1.2~GHz for the proton resonance)~\cite{Patching2025}, the steadily increasing computing power, complemented by the development of high-frequency and high-power electronics. 
In a typical solid-state (ssNMR) physics experiment, temperature, pressure, and magnetic field are essential thermodynamic variables. The accessible temperature range extends from a few dozen millikelvins to hundreds of degrees Celsius; pressures range from negative ones (using strain gauges) to over 100 GPa (using diamond anvil cells)~\cite{Semenok2026}.
The homogeneity and stability requirements of the magnetic field are much less than those mentioned above, so general-purpose high-field resistive magnets, available at several dedicated facilities worldwide, can often be used up to 35~T, or even 45~T in hybrid magnets. 
In this case, high magnetic fields are not used to increase the sensitivity or resolution of NMR spectroscopy, but rather as a physical variable that can induce phase transitions, even at zero temperature (so-called quantum phase transitions (QPT)), and to access new (quantum) phases of condensed matter.

In this review paper, we will show the interest of performing NMR in ultra-high magnetic fields to explore new field-induced quantum ground states in condensed matter. We will limit ourselves to quantum magnets, but many other fields could be considered, such as high-$T_c$ SC, exotic SC, heavy fermions, or Dirac electrons. In general, applying a magnetic field to a quantum spin system (QSS) can lead to the emergence of novel quantum phases, and until recently, NMR has been the only technique capable of probing their microscopic structure and excitations at fields above 17~T.
This situation is now evolving, with new opportunities for X-ray measurements in pulsed magnetic fields up to 30~T~\cite{Xraypulsedfields,2023Nojiri_30T} or even 50~T~\cite{XRD_55T_2025} and for neutron scattering in static fields up to 27~T~\cite{Berlin,Pregelj2019,Fogh2024} and in pulsed fields up to 40~T~\cite{Duc2018,Knafo2016}. 
Correlating results from these methods with NMR data opens a new and fascinating research direction.

For the continued advancement of NMR spectroscopy, a central objective is the realization of progressively higher magnetic fields. The strongest fields currently employed in NMR are pulsed, enabling experiments at fields approaching 100~T with durations of only a few milliseconds. In the following text, we address the prospects and some of the challenges associated with such experiments. Although the principles of pulsed-field NMR (PFNMR), which were recently reviewed in detail in Refs~\cite{Liu2021,2024_KuehneIhara} and are illustrated by several examples below, are broadly applicable to a wide range of NMR techniques, our discussion will primarily focus on solid-state systems in condensed matter physics, even more narrowly on QSS, where PFNMR spectroscopy is most commonly used.

One purpose of this review is to show how important the magnetic field is in the physics of QSS, and to concentrate on their microscopic properties and low-energy excitations.  We review several NMR-based contributions to high magnetic field physics, including those carried out by the authors using resistive magnets at LNCMI, the Laboratory National Des Champs Magnétiques Intenses in Grenoble (EMFL-G), HFML-FELIX institute in Nijmegen (EMFL-N), and at the National High Magnetic Field Laboratory (NHMFL) in Tallahassee, Florida (USA). We also present recent NMR results obtained in pulsed magnetic fields of up to 33~T at the ISSP, Institute for Solid State Physics in Kashiwa; 54~T at the HLD, High Magnetic Field Laboratory Dresden (EMFL-D); and 55~T at the LNCMI-Toulouse (EMFL-T).

\section{\label{NMRobs}NMR observables}
\label{NMR} 
Several excellent textbooks provide comprehensive introductions to the concepts and applications of NMR \cite{Abragam1961,Fukushima1981,Mehring1983,Slichter1990}. Without going into the experimental details of NMR~\cite{Fukushima1981,NarathNMR,Arcon2017}, we will restrict ourselves to its basic principles in order to clarify which physical quantities can be measured. In a most typical
case, NMR relies on the Zeeman interaction
\begin{equation}
\mathcal{H}_{\mathrm{Z}}= - \bf{\mu}_{\mathrm{n}}%
\cdot\mathbf{H}_{\mathrm{n}}
\end{equation} 
of the magnetic moment of nuclei (of selected atomic species) 
$\mu_{\mathrm{n}}=\hbar \gamma _{\mathrm{n}}\mathbf{I}_{\mathrm{n}}$, 
where $\gamma _{\mathrm{n}}$
and $\mathbf{I}_{\mathrm{n}}$ are the gyromagnetic ratio and
the spin 
of the nucleus, respectively, 
used to obtain information on the local magnetic field value $\mathbf{H}_{\mathrm{n}}$ at this position. 
A resonance signal is observed at the Larmor frequency corresponding to transitions
between adjacent Zeeman energy levels 
$\omega _{\mathrm{NMR}}=\gamma _{\mathrm{n}}H_{\mathrm{n}}$, 
allowing very precise determination of $\mathbf{H}_{\mathrm{n}}$, and therefore of the local, induced, 
so-called ``hyperfine field'' $\mathbf{H}_{\mathrm{hf}}=\mathbf{H}_{\mathrm{n}}-\mathbf{H}_{0}$\ 
(as $\gamma _{\mathrm{n}}$ is known from calibration on a convenient reference sample, and $\mathbf{H}_{0}$ is the external, applied magnetic field). The hyperfine field generated by the electrons surrounding a given nuclear site serves as a fingerprint of the local electronic environment. In addition, nuclei with spin $I$ > 1/2 have a non-spherical charge distribution and carry a quadrupole moment that couples to the electric field gradient (EFG) tensor created by the surrounding electronic and ionic charges. In a single crystal, the single NMR line associated with the Zeeman interaction is therefore split into 2$I$ lines, enabling a precise determination of the EFG tensor, which is highly sensitive to structural distortions or phase transitions, as well as to modulations of the electronic charge density, such as those occurring in charge-density wave (CDW) systems~\cite{Berthier78}. While NMR spectra reflect the static (on the NMR timescale) values of the hyperfine field and the EFG, fluctuations of these quantities give rise to the spin-lattice relaxation rate ($T^{-1}_1$), which probes their spectral density at the Larmor frequency.
\subsection{\label{StaticNMR}Local static NMR observables}

The general Hamiltonian for a nucleus with spin 
in a solid placed in an external magnetic field $H_{\mathrm{0}}$ has been analyzed in detail in many cases, most recently in Ref.~\cite{2017_Berthier}.

In the absence of unpaired electrons in the system,
$\mathcal{H}_{\mathrm{hf}}$ resumes to the so-called chemical shift~\cite{Abragam1961}, a tensor usually neglected in most of the metallic and
magnetic systems. In all other cases, the hyperfine Hamiltonian is dominated by the dipolar coupling with
unpaired electrons, which for one electron $\mathbf{s}$ at a
distance $\mathbf{r}$ can be written as
\begin{equation}
\mathcal{H}_{\mathrm{hf}} = 2\mu_{\rm{B}}\gamma_{\mathrm{n}} \hbar
\mathbf{I}\cdot[\frac{\mathbf{l}}{r^3} - 3\frac{\mathbf{r}(\mathbf{s}\cdot\mathbf{r})}{r^5} +
\frac{8\pi}{3}\mathbf{s}\delta(\mathbf{r})].
\end{equation}

In the absence of quadrupole coupling, the frequency of a line in an NMR spectrum gives direct access to the local magnetic field at the
position of the chosen nucleus. More precisely, this field corresponds to an average of the local field on the
time scale of the data acquisition, $\sim $10--100 $\mu$s for solid-state NMR. For systems with localized electronic
spins in particular, it is easy to see that the induced field is
linearly dependent on the spin polarization of the nearest
electronic spin(s)~\cite{NarathNMR}.

One benefit of measuring the hyperfine field is to obtain the true temperature and (or) field dependence of the magnetization of the spin system, which cannot always be obtained from bulk measurements since they may be dominated by the contribution of impurities at low temperature~\cite{Mendels2000}. In addition, it is 
important to determine whether the magnetization is uniform or distributed over a commensurate or incommensurate structure.

In the case of metals, the hyperfine interaction is responsible for the Knight shift, which can be written in the general case of a transition metal as~\cite{Winter1971}:
\begin{equation}
K = K_{\mathrm{s}} + K_{\mathrm{d}} + K_{\mathrm{orb}} + K_{\mathrm{dipolar}} +  K_{\mathrm{chem}},
\end{equation}
where $K_{\mathrm{s}}$ ($K_{\mathrm{d}}$) are  proportional to the density of states at
the Fermi level of s or s-p band (d-band), respectively,
$K_{\mathrm{orb}}$ depends on the filling of the d-band and is
proportional to the Van-Vleck susceptibility, and the chemical
shift $K_{\mathrm{chem}}$ is usually negligible. $K_{\mathrm{dipolar}}$  exists only in the presence of p or d bands, and depends on the symmetry of the lattice~\cite{Winter1971}. In three-dimensional (3D) systems and in absence of phase transition, $K_{\mathrm{s}}$ and $K_{\mathrm{orb}}$ are usually $T$-independent, while $K_{\mathrm{d}}$ is usually $T$-dependent.
There are many other situations in which the Knight shift varies strongly with temperature: quasi-1D organic conductors, itinerant antiferromagnets, or ferromagnets.

The localized electronic magnetic moments in insulators also contribute to the hyperfine field, introducing a Curie-Weiss temperature dependence in the Knight shift. When long-range magnetic order sets in at low temperatures, a spontaneous hyperfine field arises in proportion to the magnetic order parameter, which can be measured by NMR and used to identify the magnetic structure.

Let us now introduce the quadrupolar interaction. Its Hamiltonian
is  the part of the electrostatic interaction between the nuclei
and the electrons, which describes the interaction of the
EFG traceless tensor $V_{\alpha\alpha}$
with the quadrupole moment $Q$ of the nuclei. It can be written in the
frame of the principal axes $X,Y,Z$ of the EFG tensor as
\begin{equation}
\mathcal{H}_{\mathrm{quad}} =
\frac{eV_{ZZ}Q}{4I(2I-1)}[3I_Z^2-I(I+1) + \frac{1}{2}\eta(I_+^2 +
I_-^2)],
\end{equation}
where the axes are chosen to satisfy $|V_{ZZ}|\geq |V_{YY}|\geq |V_{XX}|$,
which constrains the 
asymmetry parameter \\
\begin{equation}
\eta=\frac{V_{YY}-V_{XX}}{V_{ZZ}}
\end{equation}
to $0 \leq\eta\leq1$. If  the quadrupolar coupling can be treated as a
perturbation, it is more convenient to express this Hamiltonian in
the laboratory frame where the  quantization is along the direction $z$ of the
applied field. For simplicity, assuming $\eta = 0$, the Hamiltonian in the case of first order in perturbation can be rewritten
\begin{equation}
\mathcal{H}_{\mathrm{quad}} =
\frac{h\nu_Q}{6}[\frac{1}{2}(3\cos^2\theta-1)(3I_z^2-I(I+1))],
\end{equation}
where $\nu_Q = 3eQV_{ZZ}/h2I(2I-1)$, $\theta$ is the angle between
the $Z$ direction and the applied magnetic field $H_0$ and $I\geq
1$. In a single crystal, the single line whose position is defined by the
Zeeman coupling $\gamma\hbar H_0(1+K)I_z$ is split into $2I$
equidistant components, separated by $\nu_Q(3\cos^2\theta-1)/2$.  In powder samples,  these $2I$ lines are thus distributed over a frequency range spanning  $(2I-1)\nu_{\rm{Q}}$.  For half-integer spins, there is a central line (corresponding to the $(-1/2, 1/2)$ transition), which is unaffected to first order, while for integer spins, no central line remains.
 Note that nuclei can have several isotopes of different natural abundance, with different gyromagnetic ratios and different quadrupole moments. For example, a single Cu site (spin 3/2) with a specific electronic spin polarization and charge environment will give rise to six different lines: 3 for $^{63}$Cu and 3 for $^{65}$Cu. A single B site (spin also 3/2) will give rise to just 3 different lines for $^{11}$B.

In closing this discussion of the static NMR observables, it should be emphasized that, in the most general situation - when the local symmetry is lower than tetragonal or trigonal, so that $\eta \neq 0$ - the Hamiltonian assumes a form more complex than the one considered above. In such cases, it becomes an explicit function of the angles $\theta$ and $\phi$, which define the orientation of the external magnetic field with respect to the principal axes of the quadrupolar tensor. For sufficiently strong quadrupolar couplings, an accurate description of the spectra may further require fitting to the results obtained from a fully diagonalized Hamiltonian to determine $\nu_q$, $\eta$ and $K(\theta,\phi)$.

\subsection{\label{DynamicNMR}Dynamic NMR observables}
There are essentially just two dynamic quantities used in NMR applied to 
solid-state physics. These are  the
spin-lattice relaxation rate $T^{-1}_1$ , and the spin-spin relaxation
 rate $T^{-1}_2$ . 
 Since $T_2$ is the correlation time of the transverse magnetization, measuring its decay, it can reflect several different origins~\cite{Mladen_Cargese}, o its interpretation may be challenging and will not be discussed further here.
$T^{-1}_1$  and $T^{-1}_2$  are both measured using appropriate pulse sequences~\cite{Abragam1961,Fukushima1981,Mehring1983,Slichter1990,NarathNMR}. 
 Now, $T^{-1}_1$ measures the characteristic time for the
longitudinal nuclear magnetization $M_{z}$ ($z$$\parallel$$H_{0}$)
to return to its thermal equilibrium value after a perturbation,
which is usually a destruction or an inversion of $M_{z}$. It
essentially measures the weight of the spectral density at the
Larmor frequency of the time fluctuations of the local hyperfine
field or of the quadrupole couplings. In the following, only the fluctuations of the hyperfine field are considered.  In localized spin systems, the part of the hyperfine field of interest for the relaxation can be written as:
$\delta \mathbf{h}(t) = \mathbf{h}(t) - <\mathbf{h}(t)> $ where
$<\mathbf{h}(t)>$ is the time-average value at the NMR timescale. For an applied field along the $z$ direction,
\begin{equation}
T^{-1}_{1z} = \frac{1}{2}\gamma_{\textrm{n}}^2 \int_{-\infty}^{\infty}dt~
\rm{e}^{i\omega_{NMR}t}\langle \delta \mathbf{h}_{+}(t) \delta
\mathbf{h}_{-}(0)\rangle .
\end{equation}
Thus, $T^{-1}_1$  directly measures the autocorrelation function $<\mathbf{h}(t)\mathbf{h}(0)>$, which develops following a power law in the case of 1D magnets. By transforming the autocorrelation function to the dynamical susceptibility, $T^{-1}_1$  is written in a general form as
\begin{equation}
T^{-1}_1  = (2\gamma_{\textrm{n}}^2k_BT)/(\gamma_{\textrm{e}} \hbar)^2\sum (A^2\chi (q,\omega)/\omega). 
\end{equation}
In the case of simple metals with a single conduction band, the relaxation rate is often expressed as proportional to the square of the density of states at the Fermi Level $N$($E_F$)~\cite{Slichter1990,Winter1971}. Since the Knight shift is also proportional to $N$($E_F$), this leads to the famous Korringa relationship:
\begin{equation}
K^2T_1T = \frac{\hbar}{4\pi k_{\rm{B}}} (\frac{\gamma_{\textrm{e}}}{\gamma_{\textrm{n}}})^2 = S_0,
\end{equation}
where $\gamma_{\textrm{e}}$ ($\gamma_{\textrm{n}}$) are respectively the electron (nuclear) gyromagnetic ratios and  $k_{\rm{B}}$ the Boltzmann constant.
The ratio $K^2T_1T$
 is used as an indicator of the dominant fluctuations in the system, since $K$ is proportional to $\chi\rm{(}\emph{q}=0\rm{)}$, while  $T_1T$ is proportional to $\omega_n/ {\sum_q}\chi''\rm{(}\emph{q},\omega_n\rm{)}$. In absence of electron-electron interaction, $\chi''\rm{(}\emph{q},\omega_n\rm{)}$ is flat and 
\begin{equation}
\mathfrak{S} = K^2T_1T/S_0 =1. 
\end{equation}
In the presence of ferromagnetic (FM) fluctuations, $\chi''\rm{(}\emph{q},\omega_n\rm{)}$, it is peaked at $q$ = 0, so that $\mathfrak{S}$ is smaller than 1, while in the presence of antiferromagnetic (AF) fluctuations, it is larger than 1.
 The situation is more complicated in transition metals, since $s-p$ band and $d$-band have to be treated separately. Moreover,  within the $d$-band, the relaxation due to core polarization, dipolar or anisotropic interactions, and the orbital relaxation have to be treated separately~\cite{NarathNMR,Winter1971}.
\subsection{\label{LowT}Low temperatures and NMR}

The most common treatments of the experimental aspects of performing NMR at low (i.e.
cryogenic) temperatures are quite difficult to find. A good and quite recent starting point is Ref.~\cite {Arcon2017} and the references therein.
\begin{figure}[pos=tbp]
    \begin{center}
    \includegraphics[width=1\linewidth]{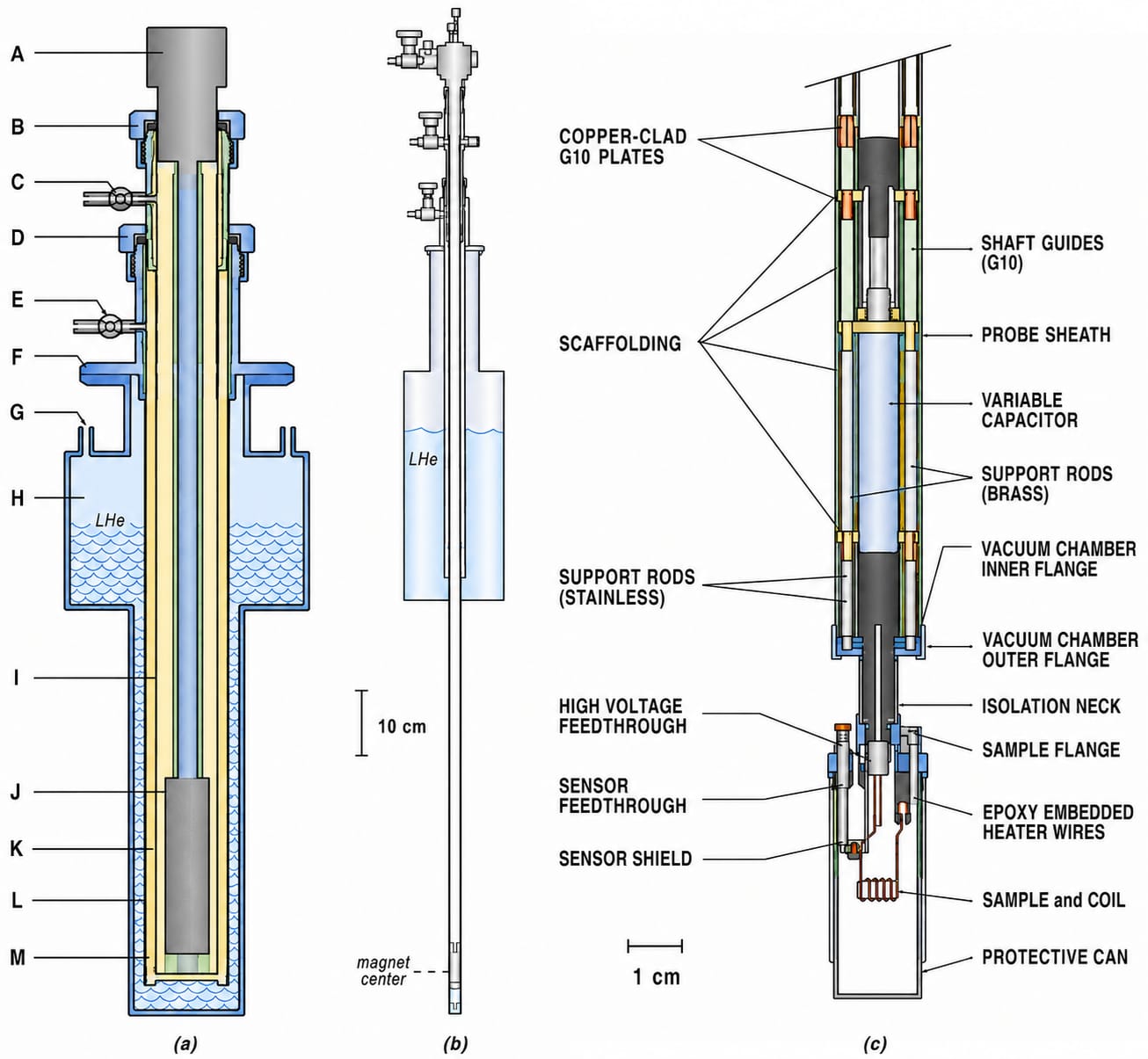}
    \caption{{\bf Example of a low-T NMR cryostat and probe}:(a) Schematic of the counterflow cryostat design. 
    A-probe head, B-quickconnect, C-exchange gas port, D-quickconnect, E-counterflow port and valve, F-Ladish flange, G-liquid He port, H-He bath, I-exchange gas space, J-probe, K-Dewar insert, L-counterflow tube, M-an annular space for counterflow. (b) Scaled drawing of the cryostat with the probe 
    inserted, emphasizing the unusual geometry of the apparatus. (c) Bottom of probe showing the details of the capacitor and its support scaffolding, the sample and coil, and temperature sensor placement. \newline Source: Adapted from Ref.~\cite{Reyes1997}.}
    \label{Fig low temp methods}
    \end{center}
\end{figure}

Performing NMR experiments at low temperatures is never a routine job and experimentalists
face several intriguing dilemmas, {\it e.g.}, those related to the choice of
cryogenic apparatus or the optimal NMR probe that could be re-tuned as the
sample temperature changes~\cite{Reyes1997,Conradi1993,Arcon2017}.
There are two typical designs of cryogenic equipment for low-temperature NMR
measurements in room-temperature bore magnets (laboratory "chemistry" SC, high-field resistive, and hybrid (resistive + SC) magnets). The first one is a so called continuous-flow cryostat
system. There, liquid He is in the He storage Dewar,
which is connected to the cryostat via a vacuum-insulated transfer tube and is transferred to the sample holder by pushing or pulling the cold He (gas or liquid) close to the sample, with the transfer stabilized by a heater/oven. Continuous-flow cryostats are popular in the magnetic resonance community because a single (liquid He) cryogen is used, the sample can be cooled in vacuum or in an exchange gas (He), they have reasonably short cool-down times of less than 1~h, and they allow for easy access to, and
replacement of the sample. Unfortunately, they do not easily allow reaching temperatures below 4.2~K. The other solution is a bath cryostat that sits atop the magnet bore, extending its long tail with the sample to the field center in the room-temperature tail (Fig.\,\ref{Fig low temp methods}). Those are often called bucket dewars and allow one to easily go below the liquid He temperature by pumping on the He bath~\cite{Reyes1997,Conradi1993}. Probes for this style can also be easily adapted for cold-bore magnets, where the temperature is regulated using their "variable temperature insert" (VTI) ~\cite{Reyes1997}. While Fig.~\ref{Fig low temp methods} presents a rather special counterflow probe design to deal with the so-called He-bubble in high-field experiments (details in~\cite{Reyes1997}), in high-field laboratories it is often preferable to have probe designs that accommodate multiple measurements~\cite{Mitrovic2022}.

\subsection{\label{HighB}High magnetic fields and NMR}

The spectral resolution in NMR depends directly on how uniform the external magnetic field H$_{0}$ is in time and space. For experiments that use NMR to determine complex molecular structures~\cite{HighRes}, the homogeneity of H$_{0}$ across a typical 1 cm sample must be of the order of 10$^{-9}$ for studies in liquid solutions, or between 10$^{-6}$ and 10$^{-8}$ for ssNMR. In both situations, the magnetic field is generated by commercially available ``high-resolution'' SC magnets that provide a {\it fixed} field. With current low-temperature SC (LTS) technology, the maximum achievable field is 23.4~T (1.0~GHz for the $^1$H resonance). Using high-temperature SC (HTS) technology, a few SC magnets with operating fields of 28.2~T~\cite{schiavina2024optimal} (1.2~GHz) and 30.5~T~\cite{wikus2022commercial,Zhang2026} (1.3~GHz) have recently become available commercially~\cite{BruWhiPap}, and Bruker is working on a 35.2~T (1.5 GHz) system. The drive toward higher magnetic fields in structural studies comes from the associated gains in both resolution and sensitivity.
\begin{figure}[pos=tbp]
    \begin{center}
    \includegraphics[width=1\linewidth]{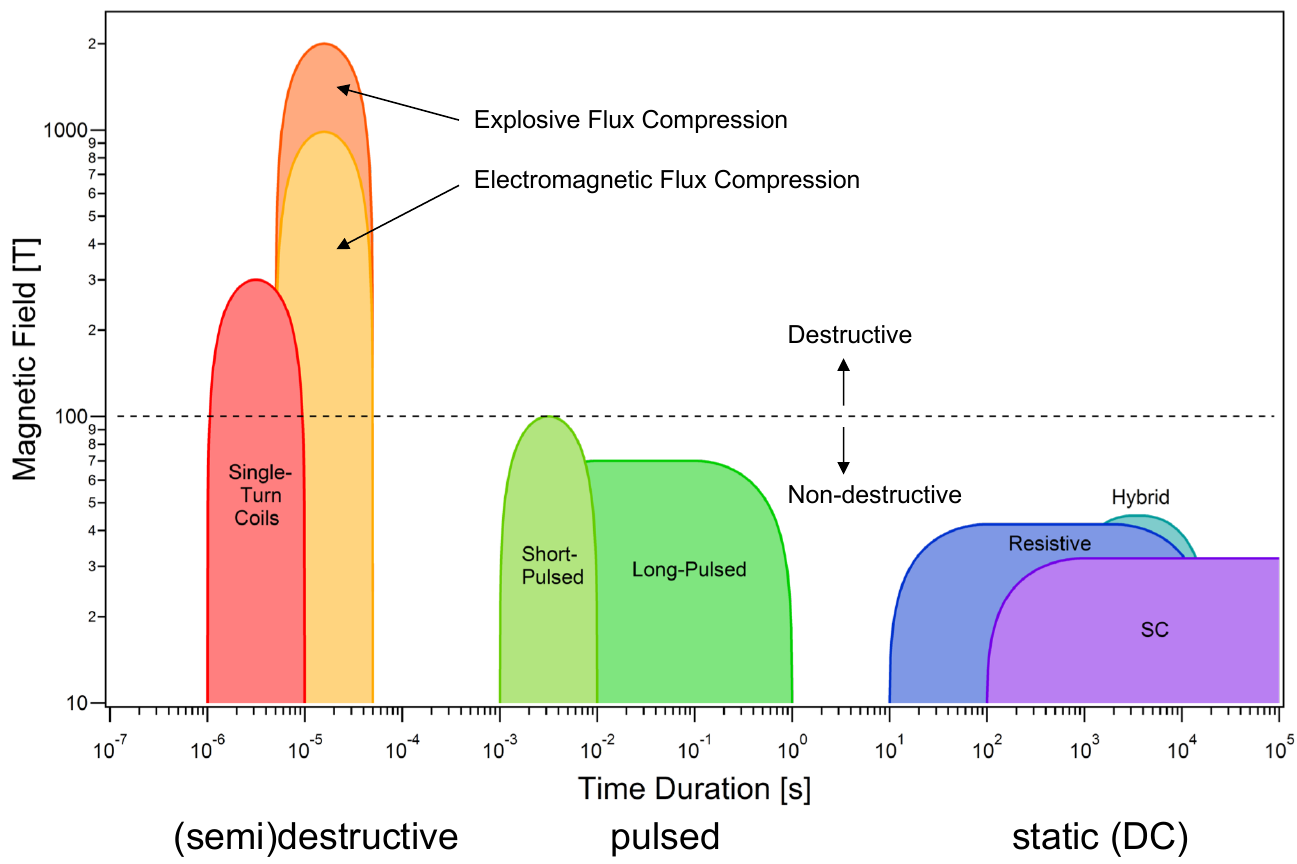}
    \caption{{\bf Overview of high-field magnet technologies}: vertical, the strength of the field; horizontal, the typical duration of the field; see also Table \ref{Table:high-field magnets}. The magnetic field values given are record values. For facility users, usually only smaller fields are available. \newline Source: Adapted from Ref.~\cite{2018_Budkin_PhysRep}.}
    \label{Fig high field magnet methods}
    \end{center}
\end{figure}

When NMR is used to investigate electronic and magnetic properties in solid-state physics, the required magnetic-field homogeneity is usually much less stringent, of the order of 10$^{-3}$--10$^{-5}$, but the field should preferably be {\it variable} (sweepable). Fields up to 22~T with this level of homogeneity can be obtained from commercial SC ``solid-state NMR magnets,'' which typically provide homogeneity of 10$^{-5}$--10$^{-6}$. Higher fields, up to about 45~T, are available from large resistive or hybrid (SC + resistive) magnets, even though their homogeneity is not optimized for high-resolution ssNMR; the available 20-25~ppm/mm is adequate for condensed-matter (typically sized) samples. Nevertheless, a typical value of 4$\times$10$^{-5}$ over a 1-2~mm sample that is accurately positioned at the field center is sufficient for most solid-state NMR experiments. Because of the small sample volume and very high operating costs, these large magnets are used mainly for NMR studies of phenomena that depend strongly on the magnetic field, such as field-induced phase transitions~\cite{PPHMF_V_p.106,PPHMF_V_p.107}. For pulsed field (PF) magnets, the same motivation holds, but for pulsed-field NMR (PFNMR), many additional challenges appear~\cite{2024_KuehneIhara,Wei2023,Wei2025}.

The approaches to addressing heat dissipation and mechanical stability challenges vary by magnet type~\cite{2018_Budkin_PhysRep,Yanagisawa_2022}, as summarized in Fig.\,\ref{Fig high field magnet methods}.
However, in all cases, materials are crucial to the development of advanced magnets; such development is thus a task at the intersection of physics, engineering, and materials science~\cite{2021_Portugall}. New developments in NMR-capable magnets are discussed further in Section~\ref{outlook}.
\begin{table}[pos=tbp]
\caption{Highest-field magnets currently in operation. The highest field does not necessarily imply a high figure of merit. EMFL facilities are labeled as D - Dresden, G - Grenoble, and T - Toulouse. Pulsed magnets are only considered nondestructive. When the record fields may originate from a single experiment, the field values available to users are significantly lower.}
\smallskip 
\centering
\setlength{\tabcolsep}{4pt}
\begin{tabular}{@{}llll@{}}
\hline \hline
Country & Field (T) & Laboratory & feature \\
\hline
\rule{0ex}{2.6ex} DC magnets &  &  &  \\

\rule{0ex}{2.6ex} USA & 45 & NHMFL & hybrid \\
\rule{0ex}{2.6ex} China & 45.2 & CHMFL & hybrid \\
\rule{0ex}{2.6ex} Europe & 42 & EMFL (G) & hybrid  \\

\hline
\rule{0ex}{2.6ex} Pulsed magnets &  &  &  \\

\rule{0ex}{2.6ex} USA & 100.7 & NHMFL &  \\
\rule{0ex}{2.6ex} Europe & 98.8 & EMFL (T and D) &  \\
\rule{0ex}{2.6ex} China & 90.6 & WHMFC &  \\
\rule{0ex}{2.6ex} Japan & 86 & IMGSL &  \\

\hline \hline
\end{tabular}
\label{Table:high-field magnets}
\end{table}

\section{\label{QSS}Quantum spin systems in magnetic field}

 Even though a spin is already by essence a quantum object, the denomination QSS is usually dedicated to systems of localized electronic spins 
 having small spin values (1/2, 1, ... , for which the eigenvalues 
 $S(S+1)$ of $\overrightarrow{S}^2$ differ significantly from $S^2$), 
 and which are dominantly coupled by AF exchange interactions. 
 In low-dimensional (1D or 2D) systems,  quantum fluctuations are enhanced, 
 and can destabilize the semi-classical long-range order (LRO) 
 ground states of N\'eel type~\cite{Auerbach,Schollwock,Giamarchi_book,Sachdev_book,Lacroix2011}. 
 Although the 1D spin chains have been studied very early by 
 theorists, they are still an important playground to study 
 experimentally modern concepts in magnetism, like, for example, 
 quantum criticality, fractional excitations (spinons), and 
 topological order. Moreover, since 1D spin systems can be 
 mapped onto 1D strongly interacting spinless fermions through 
 the Jordan-Wigner transformation~\cite{Jordan_Wigner}, they 
 allow, at the difference of other 1D systems like organic 
 conductors, nanotubes, or quantum wires, an accurate 
 verification of the predictions of the Tomonaga--Luttinger 
 liquid (TLL) \footnote{In a TLL,  all the correlation functions 
 decay as power laws. A TLL of spinless fermions is 
 characterised by two parameters $u_{TLL}$ and $K_{TLL}$, where $u_{TLL}$ is the velocity of the excitations, and $K_{TLL}$ a number which allows to calculate the exponents associated with the various correlation functions.} model~\cite{Giamarchi_book}, starting from the microscopic Hamiltonian~\cite{Klanjsek_2008}.  Concerning quasi-2D spin systems,  high \Tc~superconductors  have promoted the search for exotic quantum ground states, notably after the suggestion by P.W. Anderson that they could be described as ``a resonant valence bond (RVB) spin liquid" doped by holes or electrons~\cite{Anderson_1987}.

Several techniques, such as neutron scattering~\cite{Broholm2001}, EPR~\cite{Motokawa,Zvyagin2006}, and NMR~\cite{Mladen_Cargese}, can provide access to the microscopic properties of these new states, with neutron scattering being the most powerful.  
This section is limited to NMR in QSS under high magnetic fields and to experiments on systems in which the magnetic field plays a key role, either as a parameter of the phase diagram or of spin dynamics. This will cover NMR experiments
in quasi-1D spin chains and spin ladders~\cite{Schollwock,Giamarchi_book}, magnetization plateaus~\cite{Takigawa_11} and Bose--Einstein condensation (BEC) off triplons in coupled dimer systems~\cite{Giamarchi_natphys,Zapf2014}.

The general Hamiltonian of QSS  can be written as
\begin{equation}
\mathcal{H} = \sum_{i,j} \overrightarrow{S}_{i} \mathbf{J}_{ij}
\overrightarrow{S}_{j}  +\mathcal{H}_{\textrm{pert}}, 
\end{equation}
where $\mathbf{J}_{ij}$, the  exchange interaction (in most of the cases AF) between spins $\overrightarrow{S}_{i}$, is taken to be symmetric and can thus be written as
 $\mathbf{J}_{ij}=\sum_{\alpha} J_{ij}^{\alpha \alpha}$. The perturbation Hamiltonian $\mathcal{H}_{\textrm{pert}}$ can correspond
to the Dzyaloshinskii-Moriya (DM) interaction~\cite{Dzyaloshinskii,Moriya1956a,Moriya1956b,Moriya1960} or staggered $g$-tensors, which both correspond to antisymmetric operators mixing the eigenstates of $S^2$, to the spin-phonon interaction, and/or to the spin-spin dipolar interaction. In addition to the semi-classical ground states of
N\'eel type, two categories of ground states have to be distinguished. The basic element of both of them is the so-called
valence bond (VB) state, that is, a pair of spin 1/2 forming a singlet state. The first family is formed by  systems that can be separated
in two distinct sublattices (bipartite systems). The ground state is usually a VB Crystal (VBC), in which there is no LRO for the spins, but there is one for the VBs. Typical cases are spin ladders with strong rung couplings, and more generally, weakly
coupled spin dimer systems. 
The second one gathers geometrically frustrated systems, the archetype of which is the 2D Kagom\'e lattice, which, in some cases, are expected to have a spin-liquid ground state~\cite{Lacroix2011,Balents}. In such a state, there is no longer any LRO, neither for the spins nor for the VBs, but the ground state is a quantum superposition of many VB configurations. In this section we will mostly consider the systems in which the ground state in zero field is a collective singlet state, separated from the first excited triplet states by an energy gap $\Delta(0)$ (at $H = 0$) which can be closed at a magnetic field \Hc $= \Delta(0)/g\mu_{\textrm{B}}$.  We shall distinguish quasi-1D systems, like spin chains and spin ladders, which by using the Jordan-Wigner transformation~\cite{Jordan_Wigner} can be mapped on interacting spinless fermions and,  in most of the cases,  described in the TLL framework; and quasi-2D or 3D weakly coupled dimers in
which the triplet excitations are rather described as hardcore bosons (or "triplons"), for which the applied magnetic field plays the role of the bare chemical potential~\cite{Giamarchi_natphys}. These triplons can undergo a BEC~\cite{Giamarchi_natphys,GiamTsvelik1999,Nikuni2000} when their kinetic energy dominates their repulsion. In the inverse situation, when their hopping is strongly hindered, the triplons rather crystallize into magnetization plateaus for commensurate values of
the triplet density~\cite{Takigawa_11,Rice2002}.

\subsection{\label{BEC}Bose--Einstein condensation of triplet excitations in coupled dimer systems}
An important class of quantum spin systems can be viewed as a collection of 
dimers --~pairs of 1/2 spins, strongly coupled by an AF exchange coupling $J$ -- on a quasi-1D, quasi-2D, or 3D lattice,  coupled by weaker interdimer interactions $J'$. Such systems have a collective singlet ground state separated by a gap $\Delta$ from triplet excited states, with $ \Delta$ determined by a combination of $J$ and the interdimer $J'$. These excitations, often called triplons, can be treated as hard-core bosons on a lattice. At $T = 0$, their density is zero below the critical field \Hc and, above \Hc, is controlled by the applied magnetic field, which plays the role of a chemical potential. The hopping between neighboring sites is controlled by $J'$.
\begin{figure*}[pos=tp]
\centering
\includegraphics[width=1 \linewidth]{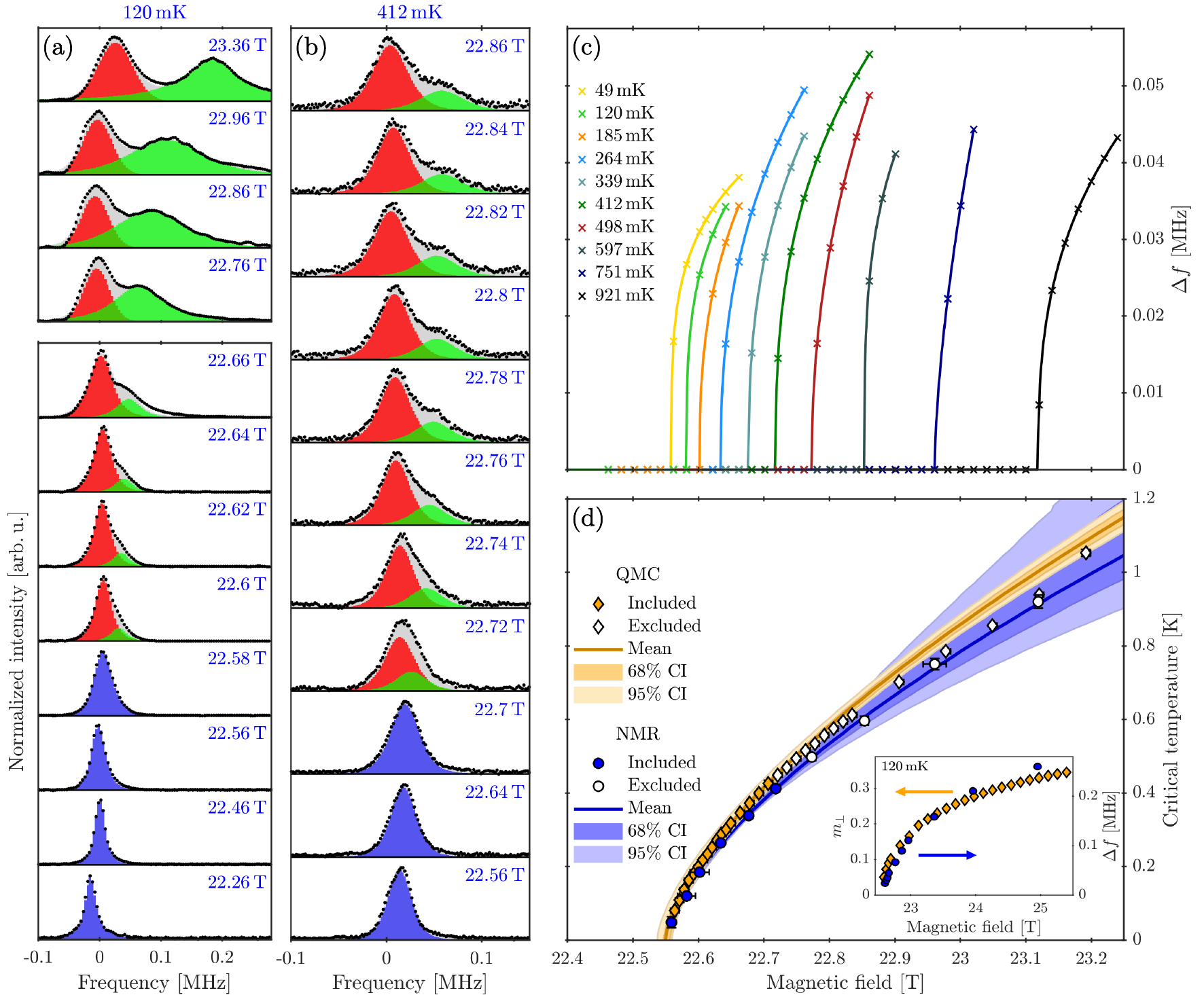}
\caption{{\bf NMR measurements of Ba$_{0.9}$Sr$_{0.1}$CuSi$_{2}$O$_{6}$ in the critical regime.} (a)-(b) $^{29}$Si NMR 
spectra measured for a range of field values around the corresponding value 
of $H_c(T)$ at 120~mK (a) and 412~mK (b). The spectra show a splitting from 
a single, sharp peak (blue) in the quantum disordered phase at $H < H_{c}(T)$ 
to two generally broader peaks (red and green) in the magnetically ordered 
phase at $H > H_{c}(T)$. The measured intensities were modeled as the sum 
(grey) of the single or double peak(s) and a weak, constant contribution. 
The double peaks were modeled by a simultaneous fitting procedure in all 
panels other than the highest four fields shown at 120~mK (a), where 
individual peak fits were used. (c) Splitting of peaks in the NMR spectra. 
Solid lines show splitting functions deduced from simultaneous fits to all 
spectra measured at the same temperature, and crosses mark the field values 
at which data were taken. (d) Low-field phase boundary of 
Ba$_{0.9}$Sr$_{0.1}$CuSi$_2$O$_{6}$ extracted from the NMR peak-splitting data 
(blue and white circles). Orange and white squares show the phase boundary 
obtained from QMC simulations. Blue and orange lines show the optimal fit 
to the phase boundary based on each set of points, obtained from the mean 
of the posterior distribution determined by a Bayesian inference procedure. 
The blue and orange shaded areas 
represent the estimated uncertainties in the form of the 68\% (dark) and 
95\% (light shading) credible intervals. NMR and QMC data points 
marked by white circles and squares are those excluded from the statistical 
analysis of the quantum critical properties. Inset: NMR peak-splitting 
measured up to 25~T at $T =~120$~mK (blue circles) juxtaposed with the 
magnetic order parameter, $m_\perp (T)$, obtained from QMC simulations.
Adapted from Ref.~\cite{Allenspach}.}
\label{fig:bscso}
\end{figure*}
Although such a description was first used to describe the superfluid properties of Helium~\cite{Matsubara1956}, it was only in 1999-2000 that the search for Bose--Einstein condensates (BEC) \cite{Bose1924,Einstein1924}  started in quantum antiferromagnets~\cite{GiamTsvelik1999,Nikuni2000}, opening a wide area of research~\cite{Giamarchi_natphys,Zapf2014}. We shall not go into the details of that field here (relevant references can be found in~\cite{Giamarchi_natphys,Zapf2014}) but rather concentrate on the NMR point of view. The condition to obtain a BEC in a QSS is the invariance of the spin Hamiltonian under a rotation around the applied magnetic field. At the onset of the BEC, a transverse ($\bot H_0$) staggered magnetization appears, which serves as the order parameter of the BEC. The transition can be observed as a function of the temperature, or as a function of the magnetic field at the quantum critical point $H_{\textrm{c}}$. For $H$ slightly larger than $H_{\textrm{c}}$ (or slightly smaller than the saturation field $H_{\textrm{s}}$), the hard-core bosons are very dilute, and one expects the transition temperature $T_{\textrm{BEC}}$ to vary as $(H-H_{\textrm{c}})^{\alpha}$. The exponent $\alpha$ is equal to $2/d$ where $d$ is the dimensionality of the system (usually $d = 3$).

From an NMR point of view, this transverse staggered magnetization splits each NMR line of the paramagnetic phase into two lines, the separation of which is proportional to the order parameter. Since NMR measures only the projection of the hyperfine field along $H_0$, the observation of this splitting requires that the hyperfine tensor components $A_{zx}, A_{zy}$ are different from zero.  Such a splitting was first observed in TlCuCl$_3$~\cite{Vyaselev2004}, but the transition at $H_{\textrm{c1}}$ was found to be a first-order one accompanied by a lattice distortion. Better examples can be found in  spin ladders compounds~\cite{Klanjsek_2008,Jeong} and in the S = 1 spin  chain NiCl$_2$-4SC(NH$_2$)$_2$ (DTN)~\cite{Zapf2006,Yin2008,Blinder2017} in which $\alpha$ = 2/3 close to \Hc and \Hs.

We shall now consider the case of the compound, known earlier as an ancient pigment Han Purple, 
BaCuSi$_2$O$_6$ \cite{Jaime2004,Sebastian2005,Jaime20267}, which has drawn a lot of interest for its peculiar low-temperature properties. 
In this quasi-2D compound, the dimers are positioned perpendicular to the $a$-$b$ plane, forming a body-centered tetragonal lattice. It was reported  that below 880~mK and close to $H=H_{c}$, $T_{\textrm{BEC}}$ was not varying as $(H-H_{c})^{2/3}$, but as $H-H_{c}$~\cite{Sebastian2006}. 
This linear dependence corresponds to a 2D BEC, and the 2D character was explained by invoking the frustration between adjacent planes due to the body-centered structure~\cite{Sebastian2006}. 
Soon after, it was shown that the system was undergoing a structural phase transition at 90~K~\cite{PPHMF_V_p.107,Samulon2006,Sparta2006,Stern2014}, giving rise to two or three alternating types of structurally~\cite{Sheptyakov2012}
and magnetically~\cite{Horvatic2005,Rueegg2007,Kraemer2007} inequivalent bilayers.

NMR experiments, performed between 13 and 26~T and at temperature as low as 50~mK,  allowed to determine the ratio of the different gaps in the two types of planes (A and B) $\Delta_\textrm{B}/\Delta_\textrm{A}$~=~1.16~\cite{Kraemer2007}, provided an accurate determination of $H_{\textrm{c}}$ = 23.4~T, and confirmed the linear dependence of $T_{\textrm{BEC}}$ with $H-H_{\textrm{c}}$ in the low temperature range. They also showed that the triplon populations of the A and B planes were very different. An accurate determination of the variation of the boson population $n_\textrm{B}$ in the plane B  with $H-H_{\textrm{cA}}$ was crucial to discriminate between different theoretical models and to explain the linear dependence of $T_{\textrm{BEC}}$~\cite{Rosch1,Rosch2,Laflorencie2009}.  Further NMR experiments have been done for this purpose in a $^{29}$Si-enriched sample. Since the average longitudinal magnetization, and hence the first moment of the NMR lines, is proportional to the boson population, it was possible to measure accurately the total boson population $n_\textrm{A} + n_\textrm{B}$ as a function of $H-H_{\textrm{cA}}$  from the first moment of the $^{29}$Si line, and the B planes boson population $n_\textrm{B}$  from the first moment of the $^{63}$Cu line, which is a very sensitive probe~\cite{Kraemer2013}. It was concluded that none of the models assuming perfect frustration could explain the very weak population observed in the B planes, and that the frustration between adjacent planes should be slightly relaxed.
The story could have stopped here, but DFT+U calculations of the exchange couplings~\cite{Mazurenko2014} eventually showed that the effective coupling between adjacent dimers was FM (in agreement with neutron data~\cite{Rueegg2007,Allenspach2020}), thus completely excluding the frustration and radically changing the microscopic scenario of this material. A new comprehensive theoretical description of this mysterious compound, including the (linear) dependence of $T_{\textrm{BEC}}$ on $H-H_{\textrm{c}}$, that of $n_\textrm{A}$ and $n_\textrm{B}$,  and the complete phase diagram in the $H$-$T$ plane, remains to be done.

The fact that, at room temperature, BaCuSi$_2$O$_6$ has a tetragonal structure (space group I4$_{1}$/acd) in which all dimer bilayers are equivalent, led to a search for a possible route to the ideal quasi-2D material by suppressing the orthorhombic transition.
After a number of stoichiometric substitution tests on the non-magnetic Ba site showed~\cite{Puphal2016} that even a 5\% Sr substitution is sufficient to stabilize the tetragonal structure down 
to the lowest temperatures. 
Later, even single crystals of the substituted material~\cite{Well2016}, and particularly 
large single crystals suitable for inelastic neutron scattering (INS) 
experiments were grown~\cite{Puphal2019}. In parallel with INS, we also presented a detailed study of the ground state, excitations, and field-temperature phase diagram of Ba$_{0.9}$Sr$_{0.1}$CuSi$_{2}$O$_{6}$~ \ cite {Allenspach}.
Fig.~\ref{fig:bscso} summarizes the main NMR results.
In the magnetically ordered phase above the critical field, which we denote by $H_c(T)$ at any temperature, the 
presence of a staggered magnetization, $m_\perp(T)$, transverse to the 
direction of the external field (which is applied along the $c$-axis) 
lifts the degeneracy between sites that were equivalent in the gapped dimer phase \cite{2017_Berthier,Arcon2017}. 
Their NMR spectra, therefore, show a splitting into two separate peaks. 
Figure \ref{fig:bscso}(a) shows $^{29}$Si NMR 
spectra obtained using a 13~mg single crystal of Ba$_{0.9}$Sr$_{0.1}$CuSi$_{2}$O$_{6}$ at $T~=~120$~mK over a range of static magnetic fields crossing the value of $H_{c}$ for this temperature. 
Whereas in BaCuSi$_{2}$O$_{6}$ it was possible by $^{29}$Si NMR to detect only a broad distribution due to the presence of structurally different bilayer types \cite{Horvatic2005,Kraemer2007,Stern2014}, whose local spin polarization could be determined quantitatively only by $^{63}$Cu NMR \cite{Kraemer2013}, in the $^{29}$Si spectra of the Sr-substituted variant, we observe a second peak that splits from the first one with increasing applied field, becoming a separate and clearly resolved entity in the upper panels of Fig.~\ref{fig:bscso}(a). 
To determine the field dependence of the peak-splitting, and hence of the order parameter, NMR spectra up to fields significantly higher than $H_{c}(T)$ for one example temperature, 120~mK, were measured [Fig.~\ref{fig:bscso}(a)], and the results are shown in the inset of Fig.~\ref{fig:bscso}(d). 
To determine the phase boundary for a study of its quantum critical properties, NMR spectra for a number of fixed temperatures between 49 and 921~mK were measured, over a range of fields close to $H_c(T)$ in each case, and a further set of spectra is shown for $T = 412$~mK in Fig.~\ref{fig:bscso}(b). 
Only at fields well above $H_c(T)$, it was possible to obtain reliable fits to the two clearly separated spectral peaks of each individual spectrum, and examples of these fits are shown in the upper panels of Fig.~\ref{fig:bscso}(a). 

The result of the simultaneous fit procedure is the set of fitting functions 
shown by the lines in Fig.~\ref{fig:bscso}(c), where the crosses represent the 
fields at which spectra were collected, and the lines yield a set of 
estimated critical points, $H_{c,j}$, with uncertainties $\delta H_{c,j}$. 
From all of this, and more from Ref.~\cite{Allenspach}, it can be concluded that the 
inter-bilayer coupling in Ba$_{0.9}$Sr$_{0.1}$CuSi$_{2}$O$_{6}$ is still 
too large to access Berezinskii-Kosterlitz-Thouless (BKT) physics, despite being only 1/100 of the magnon bandwidth (1/25 of the intra-bilayer interaction, a value 
that by any other measure appears worthy of the designation ``quasi-2D''). 
Nevertheless, the restoration of observable 3D quantum criticality from an 
apparently minor chemical substitution constitutes a somewhat remarkable 
consequence of structural ``disorder.'' It is intriguing that the spin dynamics of this compound are completely unaffected by its disordered structure. Furthermore, both Han purple and its Sr-doped version have very high critical fields: $H_{c1} \approx$ 23~T, and $H_{c2}$ exceeds the currently available maximum DC magnetic field of 45~T.

\begin{figure}[pos=tbp]
    \begin{center}
    \includegraphics[width=1\linewidth]{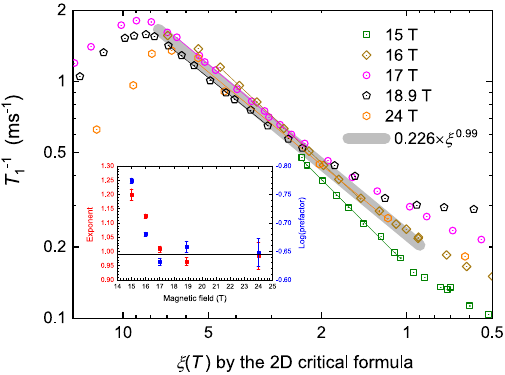}
\caption{{\bf The $^{29}$Si NMR $T^{-1}_1$ data of Ba$_2$CuSi$_2$O$_6$Cl$_2$} plotted versus $\xi$ determined by the 2D critical formula $\xi_{2D}$(T)~(Eq.~\eqref{eq8}), exhibit critical and nearly field-independent behavior in the relevant $\xi$ range (7 >$\xi$>1). The inset shows the field dependence of the parameters of the corresponding power-law fits given by the thin straight lines. The thick gray line displays the average critical behavior. \newline Source: Adapted from Ref.~\cite{Ranjith}. \textcopyright\ 2026 American Physical Society.} 
\label{fig:bscsoc}
\end{center}
\end{figure}

In contrast, the recently synthesized Ba$_2$CuSi$_2$O$_6$Cl$_2$ is a perfectly stoichiometric compound, which has the advantage that its critical fields are much lower~\cite{Okada2016}: $H_{c1} = 13.3(2)$~T and $H_{c2} = 28.5(3)$~T for the field ($H$) perpendicular to the planes, which considerably simplifies studies of its phase diagram. While the structure of the 2D planes is closely related to that of Han Purple, Ba$_2$CuSi$_2$O$_6$Cl$_2$ is not tetragonal but orthorhombic; the unit cell of a single layer contains two Cu$^{2+}$ spin dimers, but unlike in Han Purple, all the dimers are equivalent. 
The structure of the exchange couplings is a bit involved \cite{Nawa2019}, but the departure from the tetragonal symmetry is very small, so that for theoretical modeling, a simplified tetragonal model was retained: 
The spins of a dimer have the intradimer coupling $J\approx  28$~K and \emph{all} their interdimer couplings are equivalent and defined by the dominant ``diagonal'' coupling $J_d \approx 4$~K, which is slightly frustrated by the ``parallel'' coupling $J_p\ll J_d$~\cite{Okada2016}, whereas all the couplings are AF. 
An NMR study has been recently published~\cite{Ranjith}, which established Ba$_2$CuSi$_2$O$_6$Cl$_2$ as a model compound for exploring BKT criticality in quasi-2D quantum magnets. 
The NMR $T^{-1}_1$  data were used to define the BEC phase boundary and to reveal the BKT-critical fluctuations, in comparison with large-scale quantum Monte Carlo (QMC) simulations. 
The theoretically predicted 2D-critical temperature dependence of $\xi$ \cite{Kosterlitz1974,Ding1990}
\begin{equation}
\label{eq8}
\xi_{2D}(T) = \xi_0 \exp(b/\sqrt{T/T_{BKT}-1}),
\end{equation}
where the $T_{BKT}(H)$ values are fixed to those calculated by QMC, and the $b$~=~2.30 value is also determined by QMC.
 $T_1^{-1}$ obeys a simple scaling expression
\begin{equation}
\label{eq9}
T_1^{-1}(\xi) = T_0^{-1} \xi^{z-\eta},
\end{equation}
where $ T_0^{-1}$ is the power-law prefactor, $z$ is the classical exponent for the dynamics~\cite{Hohenberg1977,Borsa1992} and $\eta$ the usual BKT critical exponent. To test this scaling on their $T_1^{-1}$ data, the $\xi_{2D}(T)$ expression given by Eq.~(\ref{eq8}) was used to simply convert the $T_1^{-1}(T)$  into the $T_1^{-1}(\xi)$ representation, given in Fig.~\ref{fig:bscsoc}. Here, one indeed finds a power-law scaling behavior within the previously established range of validity of the $\xi_{2D}$ values. Furthermore, the data for all magnetic field values are nearly superimposed, meaning that $T_0^{-1}$ is practically field-independent.
Thus, the BKT transition temperature, $T_{BKT}(H)$, a key parameter in the 2D critical expression for the $T^{-1}_1(T)$ dependence, was determined, thereby allowing a precise definition of the range of applicability of 2D physics. Plotting $T^{-1}_1$  \textit{vs.} the 2D correlation length $\xi_{2D}$ provided the first experimental determination of the corresponding critical exponent, $T^{-1}_1 \propto \xi_{2D}^{1.0(1)}$~\cite{Ranjith} (see Fig.~\ref{fig:bscsoc}).

\subsection{\label{Q1D}Quasi-1D systems}
\vspace{10pt}

\subsubsection{\label{chainsladders}Spin chains and spin ladders}
{\bf Spin chains} are 1D systems made of isolated, very weakly coupled chains of strongly interacting spins. In the absence of perturbation, the spin 1/2 chains can be described by the XXZ Hamiltonian
\begin{equation}
\mathcal{H} = \sum_{i} J\{\epsilon (S_{i}^{x}S_{i+1}^{x}+%
S_{i}^{y}S_{i+1}^{y})+ S_{i}^{z}S_{i+1}^{z}\}. 
\end{equation}
For $\epsilon \geq 1$, there is no gap in the low-energy 
excitations at all values of the magnetic field and the 
spin-spin correlation functions $\langle
S_{+}(0,0)S_{-}(R,t)\rangle$ (transverse) and $\langle
S_{z}(0,0)S_{z}(R,t)\rangle$ (longitudinal) decay as power 
laws with exponents $\eta_x$ and $\eta_z$. The transverse 
correlation function, which decays more slowly than the 
longitudinal one $\eta_x <  \eta_z$, becomes dominant for 
low energy properties.  If $\epsilon < 1$, the ground state 
is of N\'{e}el type and separated from the excited states by a 
gap $\Delta(0)$, and can be closed at $H = H_{\textbf{c}}$. 
For $H \geq H_{\textbf{c}}$, the system enters into a TLL 
phase in which the longitudinal correlations are dominant 
at low energy~\cite{Haldane,Kimura_2008b}. However, 
increasing the magnetic field (i.e., the filling of the 
spinless fermions band), the exponents $\eta$ governing the 
decay of  both correlation functions cross, and above a 
second critical field $H^{*}_{c}$, the transverse one 
becomes dominant.

 Due to the inter-chain couplings  that are always present 
 in real compounds, these systems enter into a 3D ordered 
 state at finite temperature. For $\epsilon \geq 1$, 
 between $H = 0$  and the saturation field $H_{s}$,  this 
 is a canted transverse AF state, while for $\epsilon < 1$ 
 this is a longitudinal, incommensurate, spin density wave 
 between \Hc and $H^{*}$, followed by a canted AF state 
 between $H^{*}$ and the saturation field $H_{s}$.
 For example, the spin system BaCo$_2$V$_2$O$_8$  has been the 
 subject of intense investigation~\cite{Kimura_2007,Kimura_2008,Canevet_2013,Grenier_2015,Klanjsek_2015}.

In the presence of a perturbation, like an alternation of the exchange couplings, a frustrating nearest-neighbor coupling, or a spin-phonon coupling, the spin chains become gapped. As described above, applying a magnetic field can close the gap, allowing the system to enter a TLL regime. 
Another example requiring the use of a resistive magnet for NMR investigation is the orbital-assisted spin-Peierls compound NaTiSi$_2$O$_6$~\cite{NTSO}, as well as CuGeO$_3$~\cite{FagotPRL,HorvaticPRLCuGeO_99} and TiPO$_4$~\cite{TIPO}. 

Furthermore, we choose as the object of study the AF sawtooth chain~\cite{Heinze2021,Blundell2003,Moharana2026}, or delta chain, a fundamental model in 1D frustrated quantum magnetism, studied theoretically for decades~\cite{Zhitomirsky_2004,Zhitomirsky,Blundell2003}. The highest observable $^1$H resonance frequency is a key benchmark for any NMR instrument, as hydrogen is the most widely utilized NMR probe due to its abundance and strong signal intensity. Despite several instrumental challenges, it is possible to record high-quality $^1$H NMR spectra in pulsed magnetic fields with resonance frequencies up to the GHz regime~\cite{2024_KuehneIhara}. $^1$H~PFNMR measurements of the sawtooth-chain compound atacamite (Cu$_2$Cl(OH)$_3$)~\cite{Heinze2021} up to 41.2~T were crucial to probe the microscopic spin polarization across the quantum critical point at 21.9~T~\cite{Heinze2025}.
\begin{figure}[pos=tbp]
    \begin{center}
    \includegraphics[width=1\linewidth]{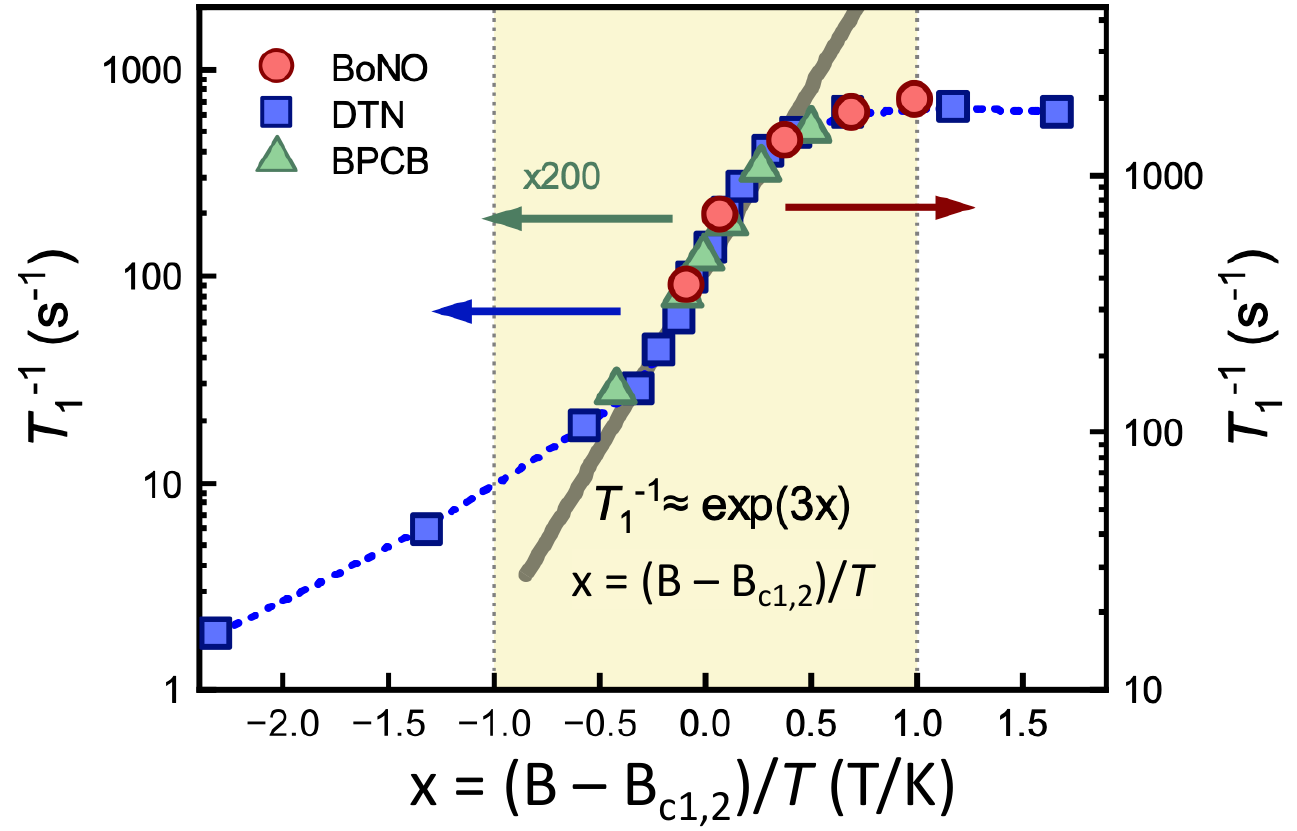}
\caption{\textbf{Spin dynamics in Tomonaga--Luttinger and quantum-critical regimes}. Universal quasiparticle scaling of $T_1^{-1}~(B)$ data measured across the quantum-critical region in BoNO (red circles), consistent with the three-magnon model (gray line), also observed~\cite{Mukhopadhyay2012} in DTN (blue squares) and BPCB (green triangles). B$_{\textrm{c1,2}}$ denotes the critical fields. \newline Source: Adapted from Ref.~\cite{Jakovac2026}.}
\label{fig:FigBONO}
    \end{center}
\end{figure}

{\bf Spin ladders} are 1D systems made of two (or more) coupled spin chains. We'll consider here only the simplest type of $S~=~1/2$ two-leg ladders, described by the following Hamiltonian:
\begin{equation}
\mathcal{H} = \sum_{i}^{l=1,2} J_{\|} \overrightarrow{S}_{i}^{l}.\overrightarrow{S}_{i+1}^{l}+J_{\bot}\sum_{i}\overrightarrow{S}_{i}^{1}.\overrightarrow{S}_{i}^{2},
 \end{equation}
where $J_{\|}$ ($J_{\bot}$) is the isotropic AF interaction along the legs (rungs). Whatever the ratio $J_{\bot}/J_{\|}$ is, the ground state of two-leg spin ladders is the collective singlet separated by a gap from the first triplet excited states. As for the spin chains, the spin ladders can be described in terms of spinless fermions. But here, the band filling starts from zero at the magnetic field \Hc, which closes the gap, and reaches complete filling at \Hs. This is different from the case of the spin chains, where the filling always starts at half-filling at zero field. At temperatures high enough to neglect the inter-ladder couplings, and low enough as compared to the Fermi energy of the spinless fermions, the low-energy excitations can be described in the framework of a TLL.
 Finding systems which are true spin ladders, with values of the AF couplings comparable to the energy scale of the Zeeman coupling for fields accessible in the laboratory, is not so easy.
Two spin ladder compounds, well characterized by NMR, neutron spectroscopy, and thermodynamic measurements, have been the subject of intense experimental and theoretical studies. The first one, (C$_{5}$H$_{12}$N)$_{2}$CuBr$_{4}$, usually called BPCB but also known as (Hpip)$_{2}$CuBr$_{4}$~\cite{Paytal,Watson}, is a strong rung coupling spin ladder  ($J_{\bot} \gg  J_{\|}$) with a \Hc~=~6.7~T and \Hs~=~11.9~T~\cite{Klanjsek_2008}. The whole phase diagram in the \textit{H}-\textit{T} plane was studied by NMR~\cite{Klanjsek_2008}, specific heat~\cite{Ruegg}, and neutron spectroscopy~\cite{Thielemann}, and the results compared to Density Matrix Renormalization Group DMRG and bosonization calculation using the values of $J_{\bot}$ and $J_{\|}$ derived from experiments.  The variation of the nuclear spin-lattice relaxation rate $T^{-1}_1$ at fixed $T$ could be remarkably fitted in the whole interval \Hc to \Hs~ using a single parameter, the hyperfine coupling. Similarly, the field dependence of the transition temperature \Tc~of the 3D phase as well as that of its order parameter could be fitted using the interchain coupling $J'$ as a single parameter. Using time-dependent (DMRG), even the high energy excitations  observed by inelastic neutron scattering could be very well reproduced, starting from the known values  $J_{\bot}$ and $J_{\|}$ of the Hamiltonian~\cite{Bouillot}.
\begin{figure*}[pos=tp]
\centering
\includegraphics[width=1 \linewidth]{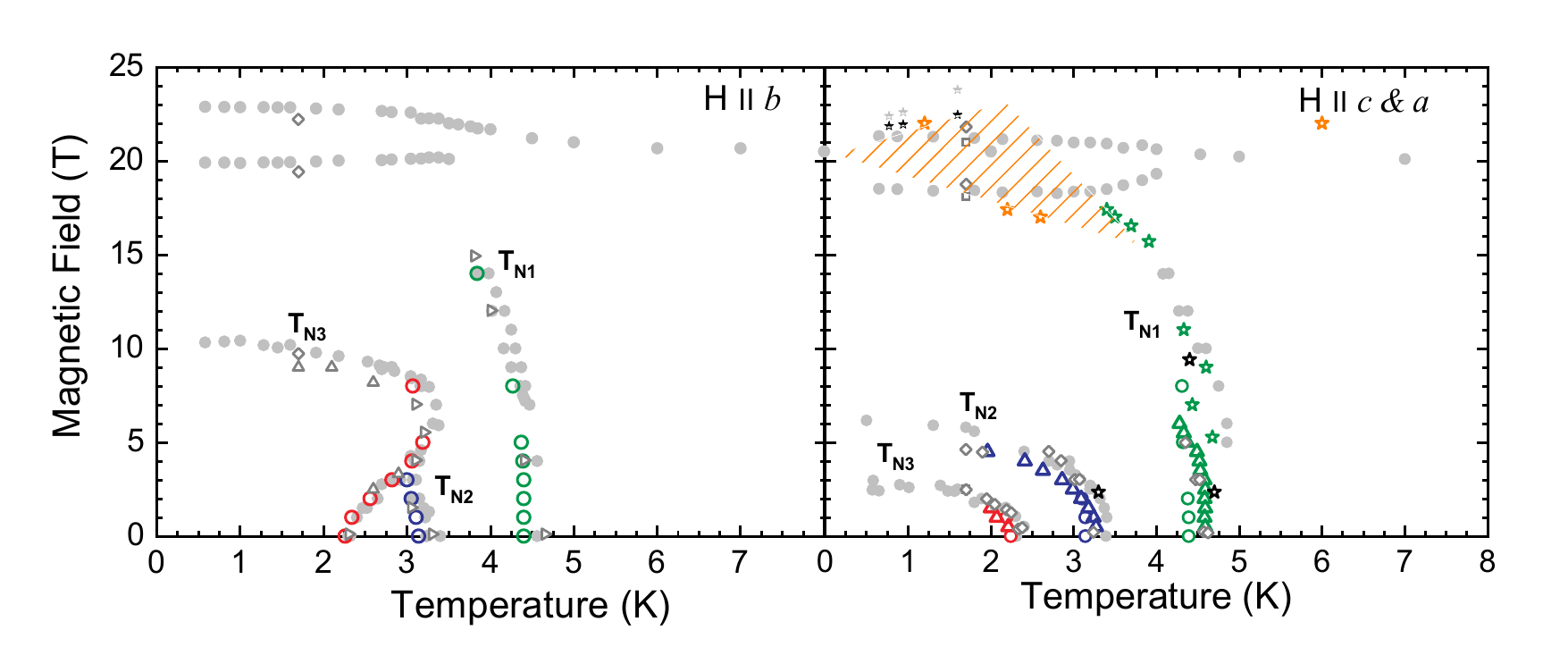}
\caption{\textbf{Field--temperature phase diagrams of $\beta$-TeVO$_4$} for magnetic fields applied along the $b$ axis (left) and along the $a$ and $c$ axes (right).
Gray filled circles are reproduced from Ref.~\cite{Weickert2016}; open squares and triangles from Ref.~\cite{Pregelj2019}; and black stars denote $^{17}$O NMR data from Refs.~\cite{Pregelj2020,Pregelj2022}.
Newly determined phase-transition data are shown as colored symbols: green ($T_{N1}$), blue ($T_{N2}$), red ($T_{N3}$), and orange (HF). Circles correspond to $C_p(T)$ measurements, triangles to $M(T)$, and stars to NMR measurements. \newline Source: Adapted from Ref.~\cite{Link2026}.}
\label{fig:PhaseDiagram_25T_included}
\end{figure*}

 The second spin ladder  compound, (C$_{7}$H$_{10}$N)$_{2}$CuBr$_{2}$ (called DIMPY)~\cite{Shapiro},  
 is in the strong-leg coupling regime 
 ( $J_{\|} > J_{\bot}$)~\cite{Hong,Schmidiger} 
 with \Hc~=~3~T and \Hs~=~29~T~\cite{Jeong2016}. Such high \Hs~=~29~T is only reachable at high-field facilities.
 This spin ladder belongs to the regime where the quasi particles are attractive \mbox{($K > 1$)}. 
 The variation of $T^{-1}_1$  as a function of $H$ between \Hc~=~3~T and \Hs~=~29~T measured at constant temperature \mbox{$T$~=~750~mK}~\cite{Jeong2016} was perfectly described by the Luttinger liquid parameters $K$($H$) and $u$($H$) determined from the starting Hamiltonian where $J_{\|}$ and $J_{\bot}$ have been determined from neutron spectroscopy. While a direct determination of the TLL parameter $KK_{TLL}$ from the expected, purely 1D power-law behaviour is rarely possible~\cite{Jeong}, allowing for the enhancement of $T^{-1}_1$ due to 3D-critical fluctuations in the random phase approximation leads to the generally applicable expression for the $T^{-1}_1(T)$ dependence that depends only on $T_c$ and $KK_{TLL}$ \cite{Orignac2020}. In this way, it is possible to determine the interaction parameter $KK_{TLL}$, and thus the nature of interactions (repulsive or attractive) in any quasi-1D spin system, even when the 3D coupling is as strong as $\simeq$10\% of the main 1D coupling.

Recently, a novel route to the design of Haldane $S=1$ chains has been taken~\cite{Jakovac2025}. 
The newly synthesized nitroxide radical-based system 3,5-bis(N-tert-butylaminoxyl)-3'-nitrobiphenyl ($m$-NO$_2$PhBNO, abbreviated as BoNO) is fully organic, with the $S=1$ units made of two ferromagnetically coupled ($|J_\text{FM}| \gtrsim 500$~K) unpaired electrons from aminoxyl groups. 
Magnetization measurements at 1.3~K show a quasi-linear increase between \Bl~$\approx 2$~T and \Bu~$\approx 33$~T, compatible with coupled $S=1$ chains with $z$\jEd /\jld = 0.04. BoNO is, therefore, a model Haldane chain system with critical field values that can not be reached in a normal laboratory setting, but are accessible at high field facilities, which allow mapping of the field-induced phases in a complete $B-T$ phase diagram~\cite{Jakovac2026}. 
The scaled data close to \Bu, where a steep $T_1 ^{-1} (B) \propto e^{-3\Delta(B)/T}$ dependence is observed, with $\Delta(B) = g \mu_B (B-\Bu)/k_B$, for BoNO lie exactly on top of the other two systems, the large-$D$ spin $S=1$ chain compound NiCl$_2$-4SC(NH$_2$)$_2$ (DTN) and the $S=1/2$ ladder system BPCB~\cite{Mukhopadhyay2012}. 

Such a behavior is driven by the critical excitations, theoretically described by three-magnon propagators~\cite{Orignac2007} that have been shown to be very robust~\cite{Ranjith2022} -- up to high temperatures. 
The exact overlap of the scaled data (Fig.~\ref{fig:FigBONO}) demonstrates that the transition at \Bu\ is a true QCP. This perfect data overlap in three different systems shows that the universality is set by an interaction-dependent scale factor, irrespective of the intrinsic nature of the system~\cite{Jakovac2026}. Finally, NMR has also proved to be a technique of choice to detect and characterize the impurity-induced quantum phases, such as the new ``order-from-disorder'' phase in Br-doped DTN compound~\cite{2017_Orlova_PRL118,2018_Orlova_PRL121}.

As a last example here, we refer to a spin-1/2 frustrated spin-ladder antiferromagnet {B}i{C}u$_2${PO}$_6$~\cite{2010Tsirlin}. Based on high-field $^{31}$P NMR experiments~\cite{Casola2013} at EMFL-G (and THz results from EMFL-N~\cite{Amelin2025}) it is argued that in this frustrated ladder compound a field-induced soliton lattice develops above a critical field of $H_{c1}$~=~20.96(7)~T (21.4~T) for field along the ladders ($b$-axis), {\it i. e.} a ladder becomes more chain-like in high fields. NMR experiments on the NHMFL hybrid also revealed a second critical field, $H_{c2}$~=~35(1)~T~\cite{Pikulski2020}, above which a field-induced chiral, or equivalently spiral, phase is established with increasing magnetic field. 

In conclusion, spin ladders and novel Haldane chains offer a rare example in which the TLL parameters can be computed directly from the Hamiltonian's microscopic parameters. In that sense, they are perfect simulators of interacting spinless fermions.

In the following sections, we discuss the frustrated $J_1-J_2$ chain $\beta$-TeVO$_4$ and the frustrated spin-1/2 chain LiCuVO$_4$, which is expected to present a spin-nematic phase at high magnetic fields around 45-- 50~T.\ \  
\subsubsection{\label{TeVO}\texorpdfstring{The frustrated $J_1-J_2$ chain: magnetic phase boundaries in $\beta$-TeVO$_4$}{The frustrated J1--J2 chain: magnetic phase boundaries in beta-TeVO4}}
Low-dimensional quantum magnets have long served as model systems for exploring the interplay between strong electronic correlations, reduced dimensionality, and magnetic frustration~\cite{Lacroix2011,Vasiliev2018}. 
In particular, one-dimensional (1D) spin chains with competing nearest-neighbor (NN) and next-nearest-neighbor (NNN) exchange interactions provide fertile ground for realizing exotic magnetic states beyond conventional long-range order~\cite{Penc2011,Starykh2015}. 
The frustrated FM--AF $J_1$--$J_2$ Heisenberg chain is especially intriguing~\cite{Zhitomirsky,Chubukov1991,Heidrich-Meisner2006,Vekua2007,Kecke2007,Sato2013,Fukuhara2013,Onishi2015,Schaefer2020,Hikihara,Sudan}, as the theory predicts a rich phase diagram that includes vector-chiral, spin-density wave and multipolar (spin-nematic) phases emerging under applied magnetic fields~\cite{Hikihara,Sudan,Nishimoto2025}. 
The vanadium tellurite $\beta$-TeVO$_4$ has been established as one of the best realizations of this model~\cite{Weickert2016,Pregelj2015,SaCrl2014}. Inelastic neutron scattering revealed the coexistence of spinon and magnon excitations in $\beta$-TeVO$_4$~\cite{Pregelj2018}, followed by the observation of a slow MHz-scale ``wiggling'' dynamics of the spin-stripe phase~\cite{Pregelj2019a}.

The high-field phase of $\beta$-TeVO$_4$ was identified by Weickert~\textit{et al.}~\cite{Weickert2016} as a potential realization of the theoretically proposed spin-nematic phase in spin $\tfrac{1}{2}$ chain systems. However, high-field neutron scattering up to 25~T by Pregelj \textit{et al.}~\cite{Pregelj2019} revealed magnetic reflections indicative of long-range incommensurate dipolar ordering, casting doubt on the presence of true spin-nematic order. With increasing magnetic field, the magnetic reflections shifted as expected for incommensurate order, with a changing SDW modulation length, but then suddenly returned near 17.2~T to values corresponding to the zero-field helical phase.
Follow-up high-field $^{17}$O NMR studies~\cite{Pregelj2020} successfully tracked the spectra of the SDW, high-field, and saturated phases. Below 1~K, the transition to the saturated phase occurred via a mixed-state regime with coexistence of spectra of the high-field and fully polarized phases. At higher temperatures, the field required for full polarization increased significantly, highlighting the role of thermal fluctuations in the high-field phase.
The saturation of magnetization observed by the NMR shift was shown to be thermally activated, thus excluding the existence of a putative spin-nematic phase. In a subsequent study~\cite{Pregelj2022}, the incommensurate high-field spectra were analyzed as two overlapping U-shaped quintets, corresponding to two neighboring magnetic chains with distinct SDW modulation axes, similar to the zero-field helical phase reported in Ref.~\cite{Pregelj2016}. 
Whereas in Weickert~\textit{et al.}~$^{125}$Te NMR~\cite{Weickert2016} was focused on the paramagnetic regime, and Pregelj~\textit{et al.}~ with $^{17}$O NMR~\cite{Pregelj2020} explored the high-field phase (20--24~T), the 2022 work ~\cite{Pregelj2022} extended the coverage to the phase transition region between the paramagnetic and ordered states.

High-field $^{125}$Te NMR measurements up to 17.4~T~\cite{Link2026} show that the resonance line shifts smoothly upon cooling in the paramagnetic regime and exhibits an abrupt drop at the onset of spin-density-wave (SDW) order, consistent with bulk magnetization. In contrast to $^{17}$O NMR, the $^{125}$Te spectrum remains a single line down to 1.4~K, reflecting the symmetric Te position within the V--O--Te--O--V super-exchange pathway. 
Spin--lattice relaxation measurements reveal that $^{125}$Te and $^{17}$O nuclei probe distinct components of the spin dynamics: Te predominantly senses fast intrachain fluctuations, while O is sensitive to slower interlayer processes. The peaks in $T^{-1}_1$  and $T^{-1}_2$ of $^{125}$Te (not shown, see~\cite{Link2026}) were used to update the magnetic phase diagram, which confirms similar magnetic behavior along the $a$- and $c$-axes (Fig.~\ref{fig:PhaseDiagram_25T_included}).

\subsubsection{\label{LCVO}\texorpdfstring{The nematic phase in frustrated $J_1-J_2$ chain: the LiCuVO$_4$ compound}{The nematic phase in frustrated J1--J2 chain: the LiCuVO4 compound}}
\begin{figure}[pos=tbp]
	\centering
	\includegraphics[width=0.95\linewidth]{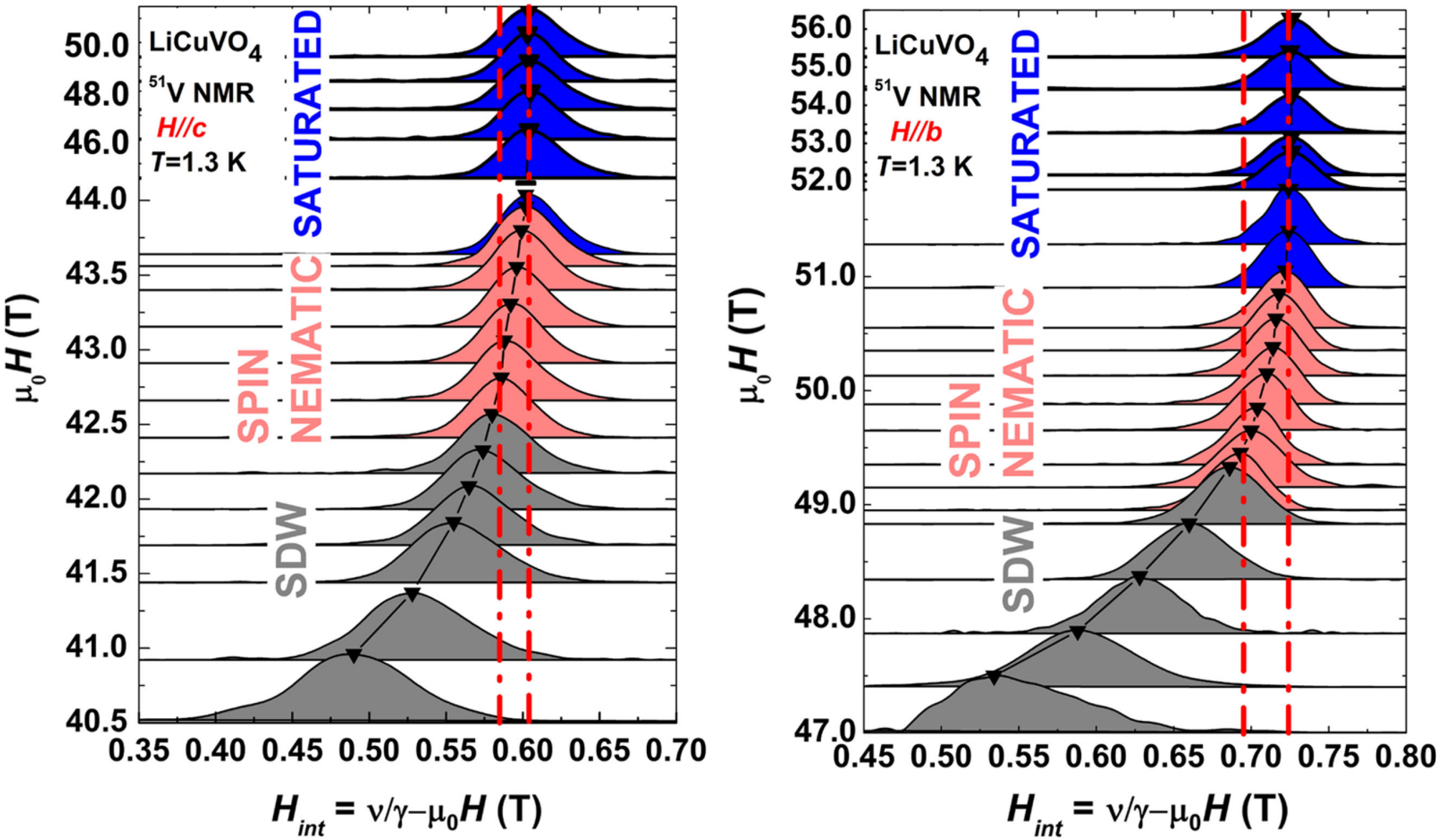}
	\caption{\textbf{Field dependence of the $^{51}$V NMR spectra of LiCuVO$_4$} for $H \parallel c$ (left) and $H \parallel b$ (right) at 1.3 K, normalized to their peak intensity. The black triangles mark the peak of each NMR line, demonstrating their shift towards the saturated state. Three different regions are identified: saturated (blue), spin-density wave (gray), and spin nematic (light red). The dashed-dotted red lines denote the region where the internal field $H_{int}$ becomes field-dependent, but maintains the same distribution as in the saturated state. \newline Source: Reproduced with permission from Orlova \textit{et al.}, Phys. Rev. Lett. \textbf{118}, 247201 (2017).  \textcopyright\ 2017 American Physical Society.}
    \label{fig:LiCuVO}
\end{figure}
In spin systems, the frustration of the exchange couplings is known to lead to exotic ground states~\cite{Lacroix2011}. Here, the case of the spin nematic phase is described, for which the compound LiCuVO$_4$ seems to be the most promising system. LiCuVO$_4$ belongs to the class of the frustrated $J_1-J_2$ spin chains, in which the first-neighbor exchange interaction $J_1$ is ferromagnetic (FM), while the next-nearest-neighbor interaction is AF~\cite{Enderle}. In such a system, the saturated FM state at \Hs~was shown to be unstable with respect to the formation of pairs of bound magnons~\cite{Zhitomirsky,Hikihara,Sudan}. In their domain of stability, these bound magnons give rise to an SDW phase, in which the transverse fluctuations are gapped, and, just before saturation, to a spin-nematic phase, in which the order parameter does not transform as a vector, but as a quadrupolar tensor of the type $S^{i}_{+}S^{i+1}_{+}$. At lower fields, single-magnon excitations prevail, giving rise to a vector chiral phase. The planar vector chiral phase and the longitudinal SDW attributed to the bound magnons have indeed been observed~\cite{Enderle,Buttgen_2012,Mourigal2012,Nawa}, and here we focus only on the nematic phase. The search for this phase in LiCuVO$_4$ was triggered by the observation of an anomaly in the magnetization curve just below the saturation field (45~T for $H\|c$, 52~T~for $H\|b$)~\cite{Svistov2011}. At such a high field, the only available \textit{microscopic} technique is NMR. The first experiment was done for $H\|c$ in a steady magnetic field~\cite{Buttgen_2014}, using the hybrid magnet at Tallahassee. It was concluded that the anomalous phase observed by magnetization was essentially due to impurities, most of the system being saturated above 41.4~T, and that the nematic phase, if present, could only exist in the range of 40.5~T~$< H < $~41.4~T. 

However, subsequent experiments conducted in the specially constructed high-homogeneity 55~T pulsed magnet at LNCMI-T~\cite{Orlova201682} for $H\|c$ and $H\|b$ have shown that, in both orientations, there is a field range below the saturation field in which there is no transverse order and the observed field dependence of the hyperfine shift is identical to the change in the magnetic susceptibility~\cite{Orlova_2017,Mila_2017}, in agreement with the theoretical predictions for a nematic phase (Fig.~\ref{fig:LiCuVO}). Further evidence could in principle be provided by measuring $T^{-1}_1$  as a function of temperature from the partially polarized phase down into the nematic one (at a fixed value of $H$)~\cite{Smerald,Orlova_2017}, but this type of measurement at such high magnetic fields is still extremely challenging.
\subsection[Multiferroics in Pb(TiO)Cu$_4$(PO$_4$)$_4$]{\label{Cupola}\texorpdfstring{Multiferroics: high-field magnetism of Pb(TiO)Cu$_4$(PO$_4$)$_4$}{Multiferroics: high-field magnetism of Pb(TiO)Cu4(PO4)4}}
The magnetoelectric (ME) response of the magnetic cluster was initially unveiled in Cr$_2$O$_3$ with antiparallel spin pairs, which are described by the sum of magnetic monopole and quadrupole moments~\cite{Folen-PRL1961}. On the NHMFL hybrid, $^{63,65}$Cu NMR on a single crystal of a multiferroic triangular antiferromagnet {C}u{CrO}$_2$ was performed with the magnetic field aligned along the crystallographic $c$-axis~\cite{Sakhratov2016}. Even though spiral 3D magnetic ordering occurs in the low-field range~\cite{Sakhratov2018}, at higher fields magnetic structures form within individual triangular planes, whereas the spin directions of the magnetic ions from neighboring planes are uncorrelated. It was established that the 2D-3D transition is hysteretic with respect to both field and temperature, and that several possible magnetic structures exist within individual planes for different phases of {C}u{CrO}$_2$. The upper boundary of the first-order transition from the 3D spiral to the 3D polar phase as the field increases at low temperatures was determined.

Subsequent studies revealed the ME response originating from combined magnetic quadrupole and toroidal moments in GaFeO$_3$ \cite{Arima-PRB2004} and further extended to those induced by the vector spin chirality of incommensurate magnetic structures in TbMnO$_3$ \cite{Kimura-Nature2003}, but the purely quadrupolar-type spin configuration had remained elusive. 
\begin{figure}[pos=tbp]
\begin{center}
\includegraphics[width=\columnwidth]{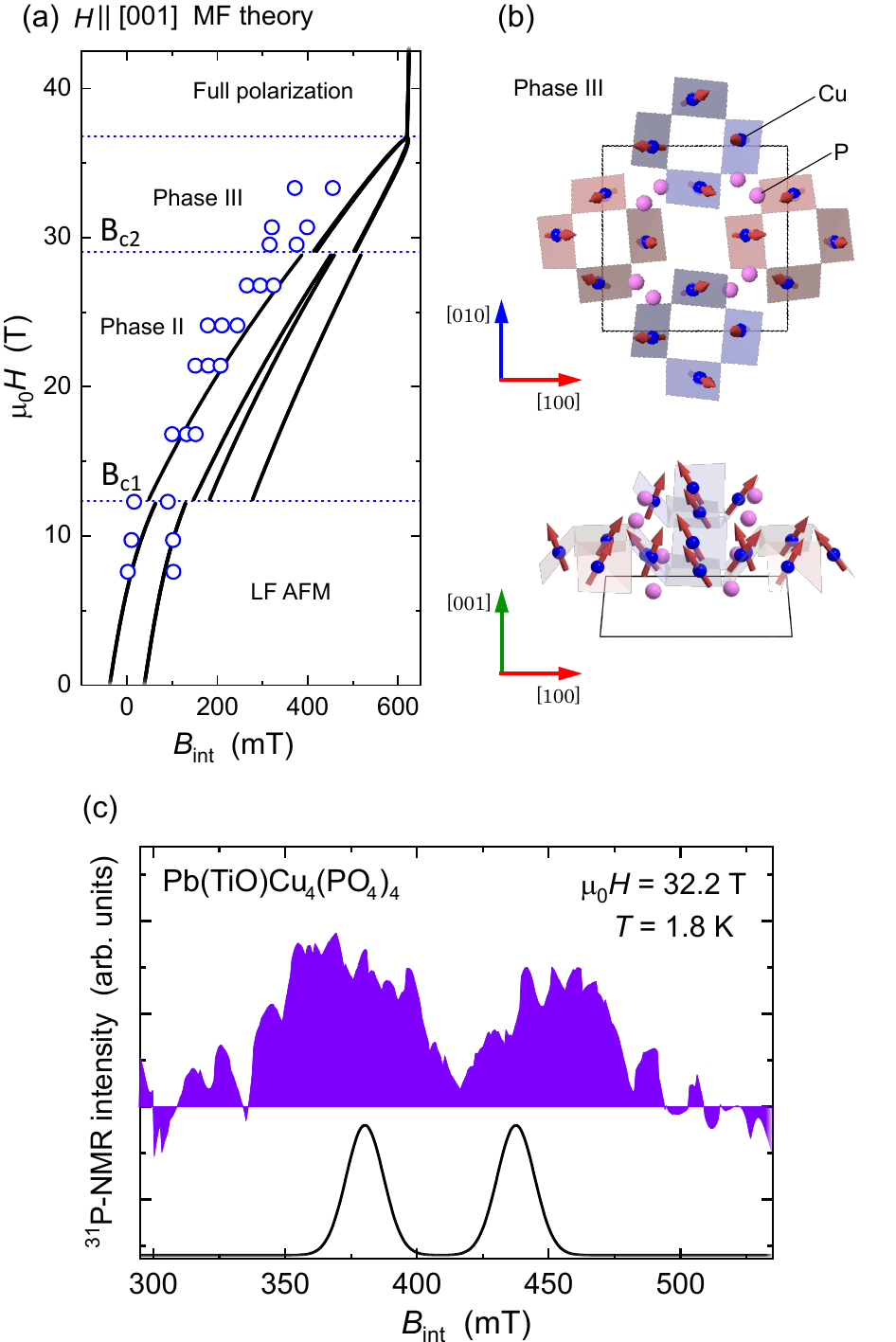}
\end{center}
\caption{\textbf{Field dependence of $^{31}$P-NMR spectra in $Pb$TCPO at low temperatures.} (a) Field dependence of $B_{\rm int}$ simulated on the basis of the magnetic structure obtained from classical mean-field theory.
Experimental data points are plotted for comparison.
Two $B_{\rm int}$ in phase III \cite{Kato-PRL2017} appearing above 29~T account for the double-peak NMR spectrum.   
(b) Magnetic structure of phase III projected on different planes. 
(c) Comparison of experimental and simulated NMR spectra.
For the simulated spectrum, a linewidth of 10~mT was introduced, and the entire spectrum was uniformly shifted by 25 \% to compensate for the anisotropy of hyperfine coupling constants~\cite{Ihara2025a}.
\textcopyright\ 2025 American Physical Society.} 
\label{fig:pIII}
\end{figure}
The magnetic quadrupolar moment was discovered in a multiferroic material $A$(TiO)Cu$_4$(PO$_4$)$_4$ ($A$ = Ba, Sr, Pb), hereafter $A$TCPO,~\cite{Kimura-NatComm2016, Kimura-PRB2018}, in which a specific spin configuration is realized on the square cupola structure of Cu$_4$O$_{12}$ clusters. A dome-like structure of Cu$_4$O$_{12}$ clusters is constructed by four walls of CuO$_4$ plaquettes (Fig.~\ref{fig:pIII} (b)). 
When the magnetic moments of Cu$^{2+}$ ions on each wall order antiferromagnetically via the dominant nearest-neighbor interaction, the two-in-two-out spin configuration is realized due to the steric angles between CuO$_4$ plaquettes and spin-orbit coupling that fixes the spin direction perpendicular to the CuO$_4$ plane.

The AF structure at zero magnetic field was solved by previous neutron and NMR measurements \cite{Kimura-NatComm2016,Babkevich-PRB96, Islam-PRB2018,Rasta-PRB2020,Rasta-PRB2026}. 
The ME response associated with the magnetic quadrupole moment was observed in $Pb$TCPO~\cite{Kimura-PRB2018}. Also in $Pb$TCPO, multiple metamagnetic transitions occur in the high-field region above 15 T, but experimental observation of the microscopic magnetic structure has never been achieved. 

NMR spectroscopy under pulsed high magnetic fields up to 32.2 T revealed the magnetic structures of the field-induced phases in $Pb$TCPO, by simulating the internal fields on the basis of cluster mean-field theory~\cite{Ihara2025a}.
Although the temporal change in the field strength imposes technical difficulty in observing the NMR signal, flat-top pulsed magnetic fields (FTPMF) extensively facilitate NMR spectrum measurement in magnetic fields even greater than 50~T \cite{2024_KuehneIhara,Ihara-RSI2021,Kohama-JAP2022}.
The NMR spectra obtained under feedback-controlled FTPMF clearly detect modifications in magnetic structures associated with the metamagnetic transitions at high fields.  
The NMR measurement at low magnetic fields was also performed to confirm the magnetic quadrupolar-type spin structure previously suggested by the neutron diffraction studies \cite{Kimura-NatComm2016}. This NMR study provides information complementary to the neutron diffraction study, namely, the local spin arrangement, including quadrupolar moments, in the Cu$_4$O$_{12}$ clusters.  

The NMR spectra in fields along the $[001]$ direction first result in double-peak spectra in the LF phase, consistent with the field-sweep spectra in steady fields.
In the intermediate fields between $B_{c1}=12.3$ T \cite{Kimura-JPSJ2019} and $26.3$ T, the peak separation is reduced, and broad spectra without clearly resolved peaks were observed. The modification of the spectral shape above $B_{c1}$ is ascribed to the metamagnetic transition already detected by the dielectric constant and magnetization measurements \cite{Kimura-JPSJ2019}.
Remarkably, another abrupt change in the NMR spectral shape was found at fields higher than $B_{c2} = 29.0$ T, which corresponds to a poorly understood dielectric anomaly \cite{Miyake-RSI2020, Nomura-PRB2023}.
The single broad spectrum splits into two peaks again above $B_{c2}$, indicating a change in the magnetic structure, even though the bulk magnetization shows only a broad hump around $B_{c2}$ \cite{Kimura-PRM2018}.  
The field variation of $B_{\rm int}$ was calculated on the basis of the obtained magnetic structures and shown in Fig.~\ref{fig:pIII}(a). 
The metamagnetic transition to phase III \cite{Kato-PRL2017} appears around 29 T, above which the double-peak structure in the NMR spectrum with two $B_{\rm int}$ is suggested. 
The calculated magnetic structure in phase III is illustrated in Fig.~\ref{fig:pIII}(b).

\subsection[Magnetization plateaus in SrCu$_2$(BO$_3$)$_2$]{\label{SCBO}\texorpdfstring{Magnetization plateaus in the quasi-2D Shastry--Sutherland compound SrCu$_{2}$(BO$_{3}$)$_{2}$}{Magnetization plateaus in the quasi-2D Shastry--Sutherland compound SrCu2(BO3)2}}
\begin{figure}[pos=tbp]
	\centering
	\includegraphics[width=0.95\linewidth]{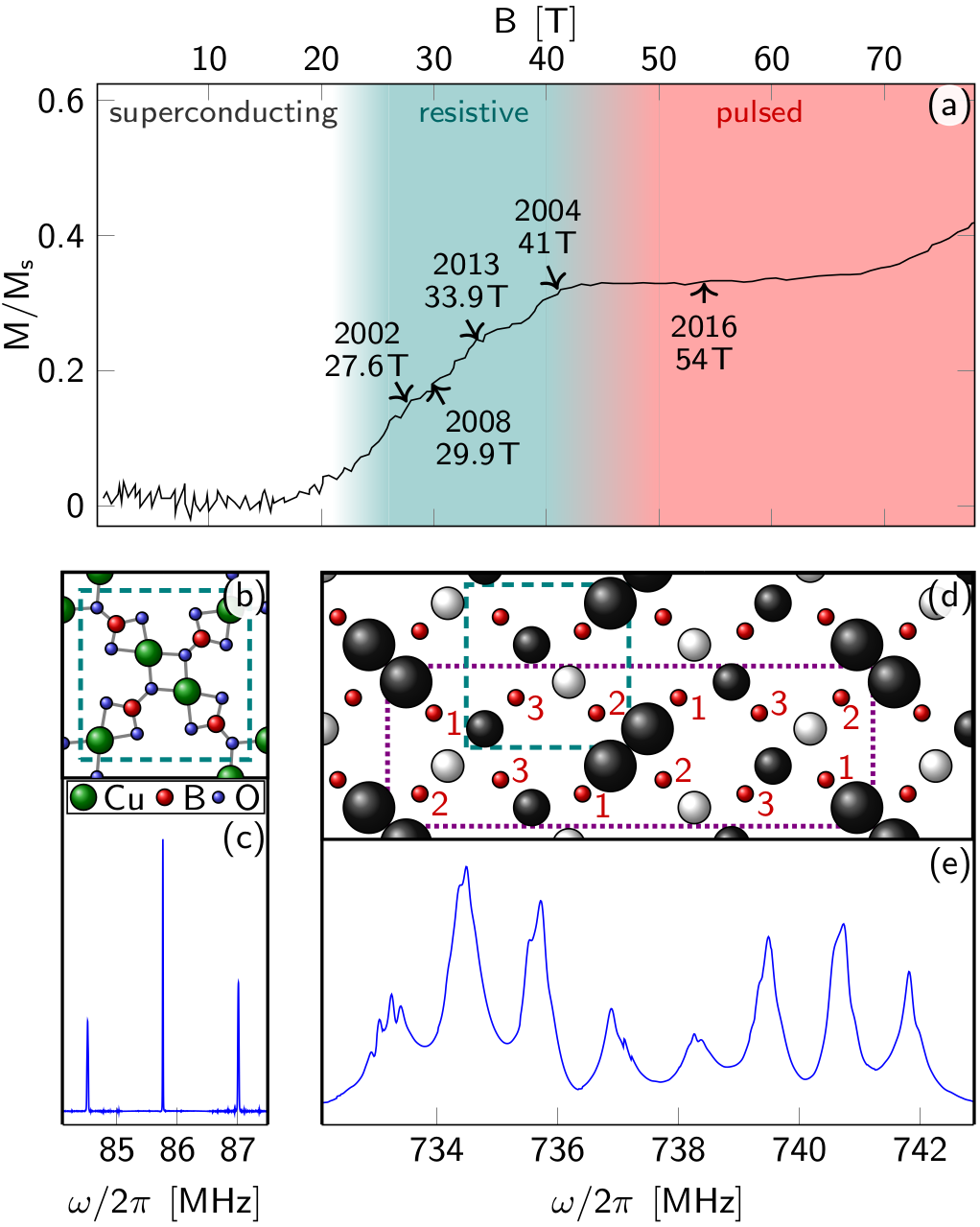}
	\caption{\label{Fig_SCBO} (a) Magnetization of SrCu$_2$(BO$_3$)$_2$ for $H \parallel c$ at 2.1 K. For previous NMR studies up to 41 T, superconducting and resistive magnets have been used (Ref. \cite{Kohlrautz2016SCBO} and citations therein). (b) Unit cell of the Cu$_2$(BO$_3$)$_2$ plane (green dashed box) and (c) the corresponding $^{11}$B NMR spectrum at 6.281 T and 5 K. (d) Magnetic superlattice at the 1/3 magnetization plateau (purple dotted box) and three inequivalent $^{11}$B sites in the plane (red spheres). White spheres and black spheres represent negative and positive electronic spin polarization, respectively. (e) $^{11}$B PFNMR spectrum at about 54 T and 2 K. \newline Source: Reproduced with permission from Kohlrautz \textit{et al.}, J. Magn. Reson. \textbf{271}, 52 (2016)\cite{Kohlrautz2016SCBO}. \textcopyright\ 2016 Elsevier.}
    \label{fig:SCBO}
\end{figure}

The Shastry--Sutherland Hamiltonian (SSH)~\cite{Shastry,shastry2026groundstateentropyisingmodel}  describes a hypothetical 2D network of orthogonal dimers, which applies surprisingly well to a real material,
the prototype
compound  SrCu$_{2}$(BO$_{3}$)$_{2}$~\cite{Kageyama1999} for the study of magnetization plateaus. In the SSH, one
considers a square lattice that is paved by  orthogonal dimers
with an AF exchange $J$ along the diagonals (next nearest
neighbors on squares, intradimer coupling), and then introduces a frustrating AF coupling $J'$
between each nearest neighbor (along sides of squares, interdimer coupling).  The product of
singlets on the dimers is always an eigenstate of the SSH, whatever the
ratio $J'/J$ is, but it remains the ground state only for $J'/J < 0.67$. For larger values of $J'/J$, the system enters a narrow plaquette
phase~\cite{Koga2000,Corboz2013} before turning to a N\'eel
ground state. In the frustrated, two-dimensional spin-1/2 SrCu$_{2}$(BO$_{3}$)$_{2}$ compound, $J'/J = 0.63$, so the ground state is the product of singlets on each dimer. However, the hopping of a triplet from one dimer to its nearest neighbor is strongly restricted,  favoring the existence of magnetization
plateaus, so a series of magnetization plateaus appears above
27~T. The first evidence for their  existence  in
SrCu$_{2}$(BO$_{3}$)$_{2}$ was obtained by magnetization
measurements in a pulsed magnetic field~\cite{Onizuka2000} and on resistive magnets~\cite{Jorge2005} with the
observation of three plateaus at 1/8, 1/4, and 1/3 of the
saturation magnetization. The first microscopic insight into the spin pattern of the 1/8 plateau was obtained soon after by Cu NMR in a resistive magnet at the LNCMI~\cite{Kodama2002,Takigawa2004}. The study of these plateaus by NMR is of high interest in order to
distinguish between different theoretical models. 
 
A single Cu site (6 lines) is observed as the magnetization grows from zero in the gapped state to the plateau, indicating uniform polarization of the Cu$^{2+}$ electronic spins. Inside the plateau, at least 10 different sites (60 lines)
 appear, spread over a distribution of internal field 
 $A_\mathrm{hyp}^\mathrm{onsite}\langle S_z\rangle$  
 of the order of 30~T. In particular, two strongly polarized sites are well resolved on the left side of the spectra (the on-site hyperfine coupling for the Cu nuclei is strongly negative, so Cu lines corresponding to strongly polarized Cu sites are strongly shifted to low frequency). Although the NMR spectra clearly demonstrated that the triplets crystallize within a commensurate supercell, they only provide a histogram of the internal field due to these crystallized triplets, not the real magnetic structure inside the unit supercell or its symmetry. Exact diagonalization of the  Heisenberg SSH on a 16-spin cluster led to the conclusion that the unit cell (16 Cu$^{2+}$ spins per layer) was rhombohedral, with oscillations of the spin polarization inside. Further calculations for the 1/4 and 1/3 plateaus showed the existence in all cases of a strongly polarized ``dressed triplet" extending on three dimers, forming stripes.
\begin{figure}[pos=tbp]
	\centering
	\includegraphics[width=1\linewidth]{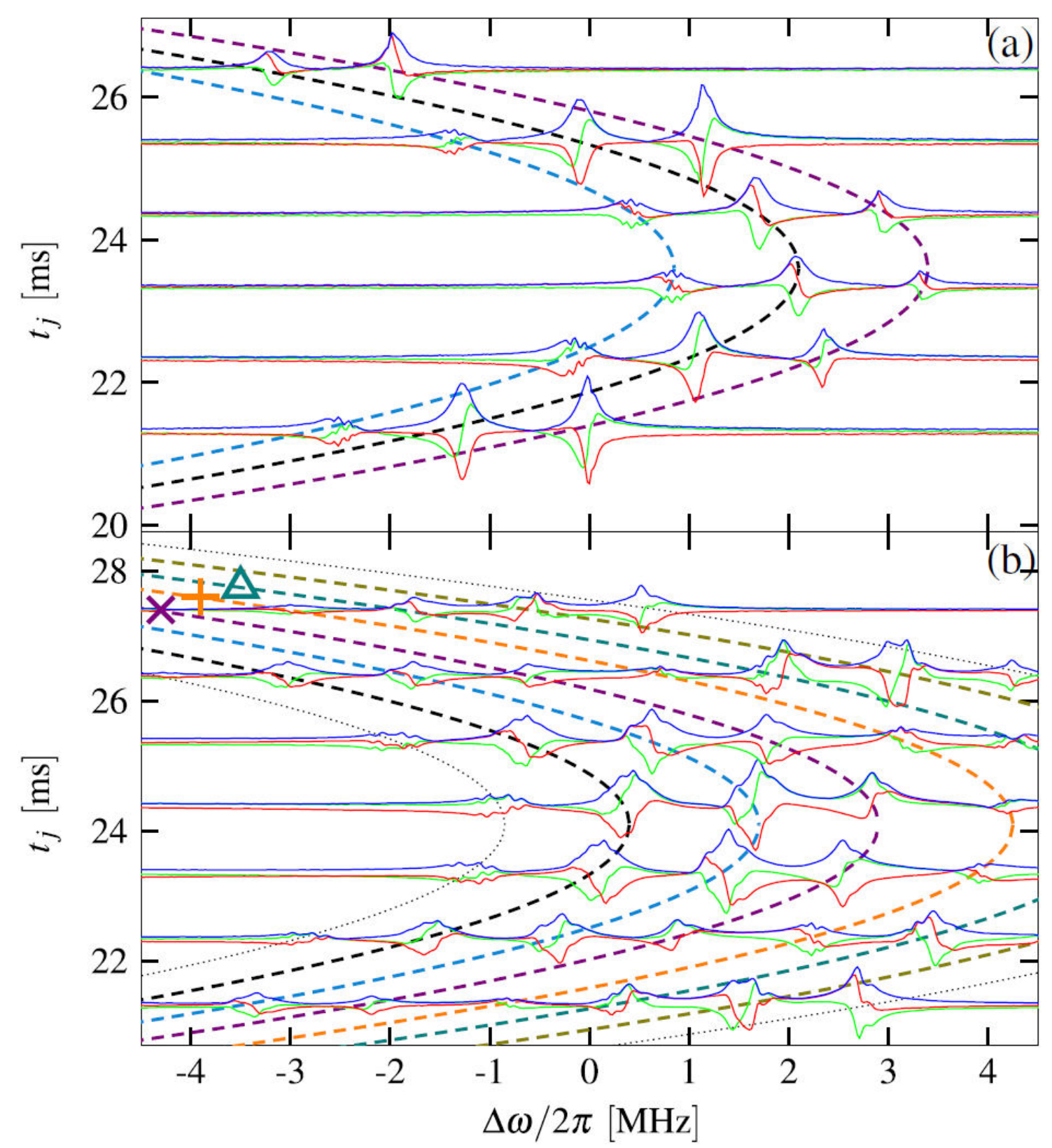}
	\caption{\label{fig:SCBO_spectra} {\bf $^{11}$B PFNMR spectra of SrCu$_2$(BO$_3$)$_2$ at 54~T} Magnitude (blue), real (red), and imaginary part (green) of the $^{11}$B spectra, recorded at different times $t_j$ at a field of about 54 T at (a) 119 K and (b) 2 K. Frequency offsets $\Delta \omega$ are defined relative to the carrier frequency of the RF pulses. Dotted/dashed curves show the approximate time-dependent field for each resonance line. \newline Source: Reproduced with permission from Kohlrautz \textit{et al.}, J. Magn. Reson. \textbf{271}, 52 (2016)~\cite{Kohlrautz2016SCBO}. \textcopyright\ 2016 Elsevier.}
\end{figure}

Further torque measurements~\cite{Levy2008,Sebastian2008} showed that the plateau sequence was not limited to 1/8, 1/4, and 1/3.
While Ref.~\cite{Levy2008} reported the existence of a second
plateau adjacent to the 1/8 plateau also observed through $^{11}$B
NMR, Ref.~\cite{Sebastian2008} claimed the existence of a full
series of plateaus at values of 
$\frac{m_z}{m_\mathrm{sat}} = 1/q$ (2~$\leq~q~\leq 9$ )  and 2/9. 
This was followed by new theoretical
attempts to calculate the sequence of plateaus and their stability~\cite{Dorier2008,Abendschein2008}, rendered difficult by the
proximity of the quantum critical point at $J'/J = 0.67$, by the necessity to take into account long-range interactions between the triplets, as well as the additional terms to the SSH Hamiltonian like DM interaction~\cite{Nemec2012}. 
With improvements in the pulse magnetic field set-ups, the 1/2 plateau was observed from 84~T~\cite{Jaime2012} to 118~T~\cite{Matsuda2013}. 
The exact sequence of plateaus up to the 1/4 plateau, that is $\frac{m_z}{m_{sat}} = $ 1/8, 2/15, 1/6 and 1/4 was finally established by $^{11}$B NMR up to 34~T~\cite{Takigawa2013}, combined with ''differential'' torque measurements. 
It was observed that within the field range corresponding to a magnetization plateau, the internal fields stay constant, as expected in a gapped phase. 
After deconvoluting the full $^{11}$B spectra to get rid of the quadrupolar splitting~\cite{Mila2013} and keeping only the central lines shifted by the internal field, structures were proposed for the above-listed plateaus~\cite{Takigawa2013}.\  Recent thermodynamic and ultrasound experiments~\cite{Imajo2022} and neutron scattering`\cite{Fogh2024} have hinted at a hidden, spin-nematic state around 26~T ''on the ground floor of the magnetic Devil's staircases".
However, a new theoretical approach (infinite projected entangled-pair states, iPEPS) finally established that the more stable structures of the plateaus consist of triplet-bound states with $S_z = 2$~\cite{Corboz2013,Corboz2014}. 
Using NMR spectra to distinguish these new structures from the previous ones is delicate because the dipolar field of the electronic spins in the adjacent planes affects the $^{11}$B spectra. 
A more direct approach would be to repeat the Cu NMR measurements in static fields with better control over line intensity. 
Another possibility is a direct neutron measurement at 26 T~\cite{Fogh2024}, provided that applying pressure would lower the threshold field of the 1/8 plateau~\cite{Schneider2016,Zayed2017} sufficiently. 

In order to probe the distribution of local spin polarization in the field range of the 1/3 plateau, $^{11}$B PFNMR was employed at fields up to 54~T and at low temperatures; see Ref.~\cite{Kohlrautz2016SCBO} and citations therein.

The broken symmetries of the plateau-specific magnetic superlattices lead to complex NMR spectra, previously investigated using steady-field NMR up to 41~T~\cite{PPHMF_V_p.106,Stern041}, which is at the lower field end of the plateau with 1/3 of the saturation magnetization, see Fig.~\ref{Fig_SCBO}(a). At this plateau, the commensurate magnetic unit cell is enlarged by a factor of\ \  3, see Fig.~\ref{Fig_SCBO}(d), and the hyperfine fields yield a spectrum that is spread over several MHz, see Fig.~\ref{Fig_SCBO}(e) and Fig.~\ref{fig:SCBO_spectra}. The $^{11}$B PFNMR FID spectra were recorded in the peak region of a pulsed magnetic field oriented parallel to the crystallographic $c$ axis. Since the field maximum was not known with sufficient precision prior to the experiment, a sequence of RF pulses with a fixed frequency of 740 MHz, a duration of 0.2 $\mu$s each, and a delay of 500 $\mu$s was employed. A pulse power of about 50 W was chosen to ensure that an on-resonance pulse would reduce the longitudinal nuclear magnetization by less than 2$\%$.

Representative results of the obtained $^{11}$B PFNMR spectra at about 54~T and at 119~K and about 2~K are shown in Fig.~\ref{fig:SCBO_spectra}. Since the $^{11}$B isotope has a nuclear spin of $I = 3/2$, the spectrum at 119 K in the paramagnetic phase (Fig.~\ref{fig:SCBO_spectra} (a))\ consists of three resonance peaks, arising from the quadrupole interaction. The time dependence of the field is indicated by the dashed curves for each resonance line. The central and satellite transitions can be identified, and a drop in intensity at high-frequency offsets is observed due to the limited excitation bandwidth of the RF pulses. 
Based on the central transition and the high-frequency satellite transition, a quadrupole splitting of 1.24 MHz is deduced, which is in good agreement with expectations from steady-field experiments (Fig.~\ref{Fig_SCBO}(c)). Furthermore, the two satellite peaks are not identical in their shapes. This can be attributed to a variation of the local electric field gradient and magnetic hyperfine shift tensors over the different $^{11}$B sites in the crystallographic unit cell of SrCu$_2$(BO$_3$)$_2$. Even a small misorientation between the axis of the pulsed magnetic field and the crystallographic $c$ axis will break the symmetry of the satellites.

\begin{figure}[pos=tbp]
	\centering
	\includegraphics[width=1.0\linewidth]{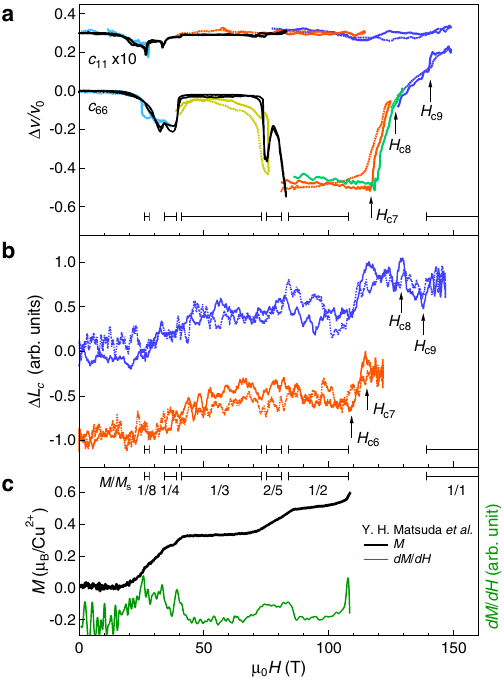}
	\caption{\label{fig:SCBO_magne}\textbf{The ultrahigh-field data ($H//c$) obtained in SrCu$_2$(BO$_3$)$_2$.}
\textbf{a} Relative change of the sound velocity for the $c_{66}$ and $c_{11}$ acoustic modes.
The results for the $c_{11}$ mode are multiplied by 10.
The results obtained with the non-destructive magnet at 1.5~K and the STCs at 3.2~K are shown by the black and colored curves, respectively.
The purple, green, orange, yellow, and cyan curves represent the results up to the maximum fields of $\sim$150, 130, 120, 70, and 30 T, respectively.
\textbf{b} Magnetostriction along the $c$ axis measured at 4.2 K.
The results up to the fields of 122 and 146~T are shown by orange and purple, respectively.
\textbf{c} Magnetization (left) and its field derivative (right) measured at 2.1 K \cite{Matsuda2013}.
The plateau field regions suggested by the magnetization measurement are shown by bars.
Anomalies above the 1/2 plateau are denoted by arrows.
The results in the field up- and down-sweeps are shown by the solid and dotted lines, respectively.
Reproduced with permission from Nomura \textit{et al.}, Nat. Comm. \textbf{14}, 3769 (2023)~\cite{Nomura2023}. \textcopyright\ 2023 The Author(s).}
\end{figure}

At 2~K, the emergence of the magnetic superstructure results in a complex distribution of static hyperfine fields, probed by NMR at the atomic $^{11}$B positions, as shown in Fig.~\ref{fig:SCBO_spectra}(b). The stacking of the Cu$_2$(BO$_3$)$_2$ layers further increases the number of inequivalent $^{11}$B sites, and the overlap of the different quadrupolar-split, site-specific peaks requires a sophisticated deconvolution procedure to elucidate the magnetic structure. The observed spectrum (Fig.~\ref{Fig_SCBO}(e)) closely resembles that previously reported at a steady field of 41~T and at 2~K~\cite{Stern041,PPHMF_V_p.106}, indicating a very similar distribution of hyperfine fields.  The most recent high-field explorations, enabled by single-turn coils, have revealed the high-field physics of this fascinating compound~\cite{Nomura2023} up to  140~T (see Fig~\ref{fig:SCBO_magne}). As the newly discovered critical fields above the 1/2 plateau suggest, there may remain a plethora of challenges and open questions for PFNMR.
These results establish the efficacy of PFNMR in studying complex magnetic order in bulk materials at the highest magnetic fields. For the very recent news on NMR of this fascinating system from the new Grenoble hybrid magnet, please check the next Section~\ref{outlook}.


\section{\label{outlook}Conclusion and outlook}
In this review paper, we have shown the interest of performing NMR in ultra-high magnetic fields to explore new field-induced quantum ground states in condensed matter. We have limited ourselves to quantum magnets, but many other fields can be considered, such as high-$T_c$ SC, exotic SC, heavy fermions, or Dirac electrons. Here are, of course, some limitations of NMR relative to other techniques: only nuclei with sufficient isotopic abundance and a suitable gyromagnetic ratio can be studied (although the former constraint can be circumvented by isotopic enrichment). On the other hand, tiny samples can be studied, which is impossible with neutron inelastic scattering. The paper focuses on NMR experiments performed in very-high-field resistive magnets, but the physics as a function of the magnetic field must be considered as a whole, and the distinction between superconducting, resistive, and pulsed magnetic fields is purely technical. In any case, experiments using the two last field sources should be carefully prepared at a lower field in superconducting magnets, which are less expensive, and for which the duration of experiments is not limited. We note that fields accessible with superconducting magnets used in solid-state physics have recently reached 35.6~T~\cite{IMR1,Liu2026} with excellent UHF NMR performed~\cite{IMR2,Han2024}. 

There is some recent good news on ultra-high-field NMR. Operations at the HFML-FELIX institute in Nijmegen (EMFL-N) are using five water-cooled resistive electromagnets that produce continuous fields up to 38~T. Recent progress has been made in developing a 45~T hybrid magnet at HFML-FELIX by reaching the outsert coil operating temperature close to liquid He.  The lab has also previously collaborated with the local NMR group of Arno Kentgens (Radboud University, Nijmegen) that is part of the National Dutch NMR network uNMR-NL~\cite{uNMR-NL} offering access to NMR instruments up to 950~MHz (Radboud University) and 1.2~GHz (Utrecht University). For example, van Bentum et al.~\cite{vanBentum2003} implemented experimental strategies for ssNMR studies in high-field Bitter and hybrid magnets, which are now used at many high-field facilities.

\begin{figure}[pos=tbp]
	\centering
	\includegraphics[width=1.0\linewidth]{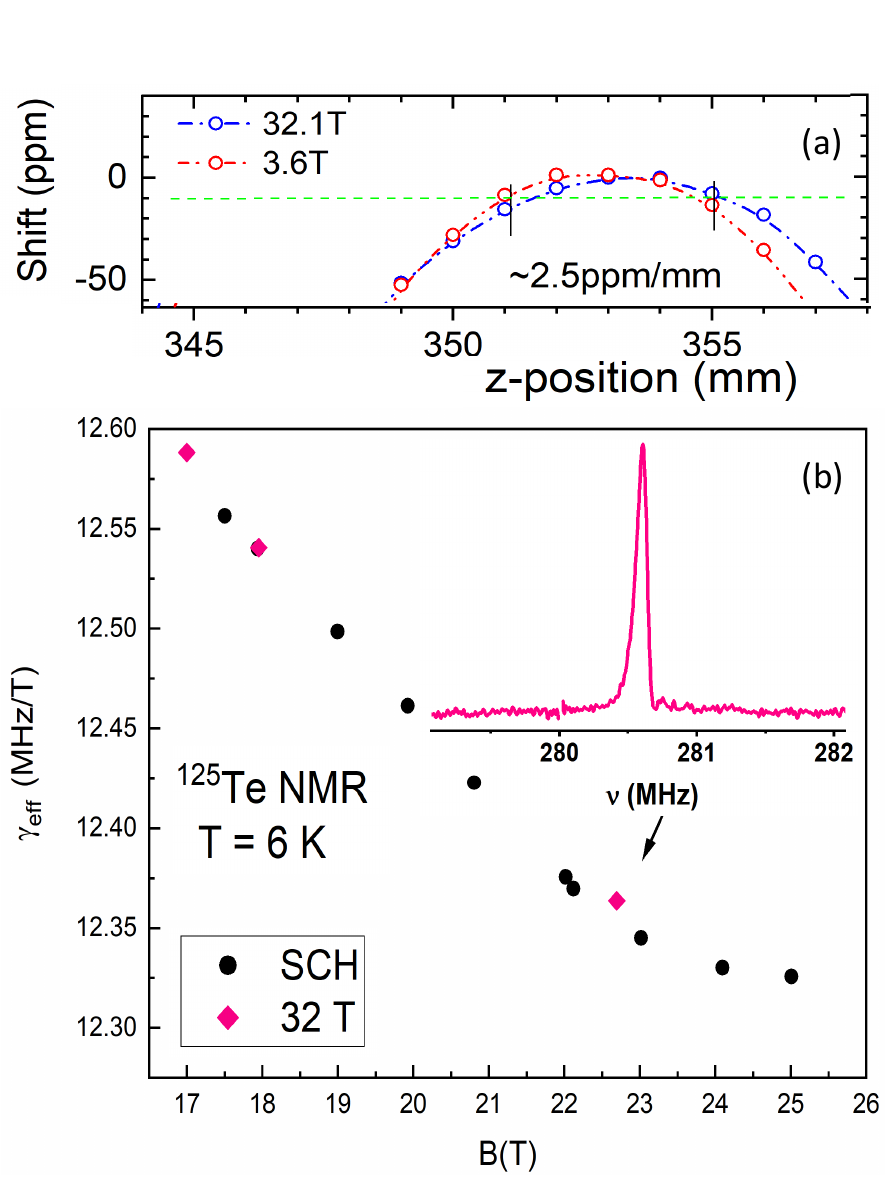}
	\caption{\label{Fig_32T} (a) Magnet field profile of the new 32~T SC magnet at NHMFL along the z-axis at 32.1~T and 3.6~T measured with $^{63}$Cu NMR~\cite{Kolb-Bond2022}. (b) $^{125}$Te spectral line and effective gamma $\gamma_{\rm eff}$(NMR frequency divided  by applied field, normal $^{125}\gamma$ ~=~ 13.454 MHz/T) with respect to applied magnetic field. Black circles correspond to the 36~T Series Connected Hybrid (SCH), while the pink diamonds correspond to the new HTS 32~T magnet.}
\end{figure}

After it reached 42~T in November 2024, some of us performed a first scientific NMR experiment at LNCMI's (EMFL-G) new hybrid magnet in June 2025~\cite{EMFLNEWS.3.25}. Two objectives were achieved during this test run, despite the short measurement period of only three days. The scientific goal was the investigation of SrCu$_2$(BO$_3$)$_2$ by $^{63,65}$Cu NMR (see above, Section~\ref{SCBO}). The first-ever broadband NMR spectrum above 40~T was recorded at 42~T and 0.4~K. The spectrum covers a range from 210 to 700 MHz and provides the local magnetic field values at the
copper sites of SrCu$_2$(BO$_3$)$_2$ in the 1/3 magnetization plateau~\cite{vinograd2026}. Prior to the scientific experiment, the field homogeneity and stability at 42~T (about 1.8~GHz for $^1$H) were recorded by $^{27}$Al NMR of a 1~mm$^3$ sample. Near the center of the magnet, the spatial field profile followed the
expected behavior and provided about 10~ppm field homogeneity in 1~mm$^3$. The standard deviation of the magnetic-field fluctuations, recorded by single-scan NMR, was 5~ppm, and the long-term drift over
6.5 hours was 10-15 ppm. These values are perfectly adequate for solid-state NMR and even slightly better than in pure resistive magnets owing to the better field homogeneity and stability of the superconducting
outsert. This study also demonstrated that the new hybrid magnet can operate at full power for more than 6.5 h. The success of this first NMR experiment marks a milestone toward regular user operation in the near future. The Grenoble Hybrid complements nicely the NHMFL existing 45~T hybrid (where the world record NMR $^1$H experiment was performed in 2001 at 44.7~T and a proton NMR frequency of 1.9~GHz~\cite{Clark2002}) and a younger Series-Connected Hybrid (SCH) magnet for NMR spectroscopy~\cite{Gan2017}, which as a DC powered magnet can be operated at fields up to 36.1~T, and enables NMR spectroscopy at $^1$H frequencies of 1.0, 1.2 and 1.5~GHz. 

High-field facility NMR users are often called "the sitters" because they must remain at high field strengths for long hours. In this context, the hybrid magnets share the same disadvantage as resistive electromagnets - high energy prices make our experiments very expensive. Here, the new developments of all-superconducting user magnets bear high hopes. 

By 2017, NHMFL had developed, built, tested, and
commissioned a 32~T all-superconducting user magnet system combining two series-connected high-field high-temperature superconductor (HTS) nested inner coils (insert) wound with SuperPower, Inc. REBCO tapes and a low-temperature superconducting (LTS) outer magnet (outsert) composed of five coils (subdivided into 17 electrical sections)~\cite{Markiewicz2012_32T}. Fig.~\ref{Fig_32T} shows results from our short run on this magnet, just after a successful use of the SCH. While the SCH demonstrates (after careful shimming) better than 1~ppm homogeneity over a cylindrical volume of
1~cm diameter and height~\cite{Gan2017}, the 32~T magnet achieves 2.5~ppm over 1~mm, clearly better than the available hybrids (Fig.~\ref{Fig_32T}(a)). Fig.~\ref{Fig_32T}(b) demonstrates the excellent agreement of $^{125}$Te shift field dependence in $\beta$-TeVO$_4$ (see above, Section~\ref{TeVO}) at 6~K. After repair, the 32~T magnet would be an invaluable tool for the low-T condensed matter NMR community. As the record field using HTS inner coils reaches 45.5~T~\cite{Hahn2019_45p5T}, the feasibility of the next 40~T all-superconducting magnet~\cite{Bai2020_40T} seems very high. Similar magnets are under construction at the EMFL-G site~\cite{Fazilleau2020,Song2022}.
To put it more specifically, the high field values, the very low noise and vibration levels, and the possibility of running very long-duration experiments will make high all-superconducting magnetic fields attractive to scientific communities far beyond NMR (scanning probes, Fourier transform infrared spectroscopy, ultra-low-temperature physics, electrochemistry, \textit{etc.}).

Obviously, the development of NMR in ultra-high magnetic fields relies on pushing this limit as high as possible. With new hybrid setups coming online, all-superconducting solutions employing high-T$_c$ SCs are catching up with the frontline of the best resistive magnets - all this field is in rapid development. Until recently, NMR was the only technique that allowed the acquisition of microscopic information above 17~T. The development of dedicated hybrid magnets for neutron and X-ray scattering under increasing pulsed magnetic fields will enable robust progress in the field through fruitful comparisons between these complementary techniques.

\section*{Declaration of competing interest}
\addcontentsline{toc}{section}{Declaration of competing interest}
The authors declare that they have no known competing financial interests or personal relationships that could have appeared to influence the work reported in this paper.

\section*{Acknowledgments}
\addcontentsline{toc}{section}{Acknowledgments}
We are indebted to all the participants to the work reported here, in particular S.~Allenspach, D.~Ar\'{c}on, M.H.~Baldus, C.~Berthier, M.~Bird, R.~Blinder, G.~Blumberg, A.~Boldin, T.~Chakrabarty, W.~Clark, Y.~Fagot-Revurat, B.~Gaulin, M.~Grbi\'c,  E.~Green, J.~Haase, W.~Halperin, M.~Hirata, M.~Jaime, M.~Jeong, T.~Kimura, M.~Klanj\v{s}ek, Y.~Kohama, J.~Kohlrautz, P.~Kuhns, J.~Link, E.~Lippmaa, V.E. Mitrovi\'c, A.~Orlova, T.~Painter, K.~Proke\v{s}, R.~R\"asta, C.P.~Slichter, M.~Takigawa, I.~Vinograd, F.~Weickert, J. Wosnitza, K.~Yoshimura, and S.~Zvyagin. We thank H.~Berger, H.~Dabkowska, H.~Kageyama, K.~Kimura, and P.~Puphal for providing excellent single crystals. We also acknowledge C.~Batista, T.~Giamarchi, N.~Laflorencie, F.~Mila, and  A.~Tsirlin for their theoretical support. RS is thankful to the US State Department Fulbright Scholar program for enabling his sabbatical at NHMFL, and to Laura H. Greene for her hospitality.

The authors acknowledge support from the ISABEL project, funded by the European Union under the Horizon 2020 Grant Agreement No. 871106. The work at NICPB was conducted using research infrastructures EstMagLab (TARISTU24-16) and AKKI (TARISTU24-15) funded by the Estonian Research Council (ETAg), and supported by the European Research Council (ERC) under the European Union Horizon 2020 research and innovation program, Grant Agreement No. 885413. The ETAg Grants IUT23-7, PRG4, PRG1702, and TemTA-25 supported this research. 
A portion of this work was performed at NHMFL, which is supported by the National Science Foundation Cooperative Agreement No. DMR-2128556 and the State of Florida. We appreciate the support of the HLD at HZDR and the LNCMI, both members of the European Magnetic Field Laboratory (EMFL), and we acknowledge COST Action 21144 (SUPERQUMAP).
Most of the work presented here was supported by the European Commission Contracts No. RITA-CT-2003-505474 (High Field research), RII3-CT-2004-506239 (EuroMag-Net), and FP7-INFRASTRUCTURES-228043 (EuroMagNet II); by the French ANR Grants No. ANR-06-BLAN-0111 (NEMSICOM), ANR-12-BS04-0012 (SUPERFIELD), and ANR-14-CE32-0018 (BOLODISS); by LANEF in Grenoble (Project No. ANR-10-LABX-0051); by the Equipex LaSUP of the Université Grenoble Alpes, CNRS and French Ministry of Higher Education and Research; by the European Funds for Regional Development (FEDER); and by the French Rhône-Alpes region.

\section*{Data availability}
\addcontentsline{toc}{section}{Data availability}
No data was used for the research described in the article.

\begingroup
\small
\setlength{\bibsep}{2pt}
\phantomsection
\addcontentsline{toc}{section}{References}
\bibliographystyle{unsrtnat}
\bibliography{Stern_SSNMR_arxiv}
\endgroup

\end{document}